\documentclass[nofootinbib,secnumarabic,amssymb,superscriptaddress, aps, pre]{revtex4}

\usepackage{amsmath}
\usepackage{mathtools}
\usepackage{graphicx}
\usepackage[dvipsnames]{xcolor}
\usepackage{epsfig}
\usepackage{mathrsfs}
\usepackage{amsbsy}
\usepackage{verbatim}

\usepackage[colorlinks,citecolor=blue]{hyperref}
\usepackage[capitalise,nameinlink]{cleveref}
\hypersetup{colorlinks=true, linkcolor=blue}
\hypersetup{urlcolor=blue}
\usepackage{capt-of}
\usepackage{cases}
\usepackage{empheq}
\usepackage{orcidlink}

\newcommand{\ba}{\begin{align}}
\newcommand{\ea}{\end{align}}
\newcommand{\nn}{\nonumber}
\newcommand{\be}{\begin{equation}}
\newcommand{\ee}{\end{equation}} 
\newcommand{\bea}{\begin{eqnarray}}
\newcommand{\eea}{\end{eqnarray}}
\def\nn{\nonumber}

\usepackage[FIGTOPCAP]{subfigure}
\usepackage[subfigure]{tocloft}
\usepackage{latexsym,oldgerm}
\usepackage{longtable}
\def\be{\begin{equation}}
\def\ee{\end{equation}}
\def\bea{\begin{eqnarray}}
\def\eea{\end{eqnarray}}
\def\nn{\nonumber}

\begin{document}
\title{General solutions of poroelastic equations with viscous stress }
\author{Moslem Moradi}
\email[Email: ]{mmoradi@email.unc.edu}
\author{Wenzheng Shi}
\email[Email: ]{swz18@live.unc.edu}
\author{Ehssan Nazockdast}
\email[Email: ]{ehssan@email.unc.edu}
\affiliation{ Department of Applied Physical Sciences, University of North Carolina at Chappel Hill, Chapel Hill, NC 27599-3250}

\date{\today}
\begin{abstract}
\vspace*{0.55cm}
\subsection*{Abstract}
\noindent 
Mechanical properties of cellular structures, including the cell cytoskeleton, are increasingly used as biomarkers for disease diagnosis and fundamental studies in cell biology. Recent experiments suggest that the cell cytoskeleton and its permeating cytosol, can be described as a poroelastic (PE) material. Biot theory is the standard model used to describe PE materials. Yet, this theory does not account for the fluid viscous stress, which can lead to inaccurate predictions of the mechanics in the dilute filamentous network of the cytoskeleton. 
Here, we adopt a two-phase model that extends Biot theory by including the fluid viscous stresses in the fluid's momentum equation. We use generalized linear viscoelastic (VE) constitutive equations to describe the permeating fluid and the network stresses and assume a constant friction coefficient that couples the fluid and network displacement fields. 
As the first step in developing a computational framework for solving the resulting equations, we derive
closed-form general solutions of the fluid and network displacement fields in spherical coordinates. 
To demonstrate the applicability of our 
results, we study the motion of a rigid sphere moving under a constant force inside a PE medium, composed of a linear elastic network and a Newtonian fluid. 
We find that the network compressibility introduces a slow relaxation of the sphere
and a non-monotonic network displacements with time along the direction of the applied force. These novel features cannot be predicted if VE constitutive equation is used for the medium.
We show that our results can be applied to particle-tracking microrheology to differentiate between PE and VE materials and to independently measure the permeability and VE properties of the fluid and the network phases. 
\end{abstract}
\maketitle


\section{Introduction}
Many biological materials, such as muscular and brain tissues, extracellular matrix and the interior cytoskeleton, are composed of  filamentous networks that are permeated by and immersed in a fluid-like phase \citep{howard2001mechanics}. The network phase provides elasticity and mechanical strength to the structure, whereas the fluid phase contains a variety of molecular components that are key to regulating different subcellular and extracellular processes. The fluid-network mechanical coupling is key to determining the deformations of the network, the resulting fluid flows and molecular transport \citep{mogilner2018intracellular}. A continuum framework for modeling of fluid flows permeating deformable structures  was developed by Biot \citep{biot1941general, biot1955theory} to study soil consolidation and other problems related 
to soil and rock mechanics, where both the fluid and the solid phases are assumed to be in their linear response regime. Aside from its application in studying geomaterials \citep{detournay1993fundamentals, cheng2016poroelasticity}, Biot theory has also been applied to studying the mechanics of biological materials including  tissues \citep{simon1992multiphase, paulsen1999computational} and bones \citep{cowin1999bone}. Theoretical and experimental studies over the past decade suggest that the cell cytoskeleton is most appropriately described as a poroelastic (PE) material, in opposed to a viscous, elastic and/or viscoelastic material \citep{mogilner2018intracellular}. As such, poroelasticity theory has been used to model the cytoskeleton mechanics \citep{moeendarbary2013cytoplasm,copos2018porous, rosenbluth2008slow,malandrino2019poroelasticity,wang2020characterizing,moradi2021cell}  and cellular processes such as cell blebbing \citep{charras2005non, charras2008life, mitchison2008implications, strychalski2015poroelastic, strychalski2016intracellular}, and cell motility \citep{strychalski2015poroelastic}. 
The governing equations of Biot's theory are \cite{doi2009gel}:
\begin{subequations} \label{eq:biot}
\begin{align}
 \mathrm{Mass \, conservation:} & \qquad \nabla \cdot \left(\phi \mathbf{v}_n+(1-\phi)\mathbf{v}_f\right)=0 ,  \\
\mathrm{Momentum\, Eq.\, for\, the\, network:}   & \qquad \nabla \cdot \left(\boldsymbol{\sigma}_n-\phi p \mathbf{I}\right)-\xi \left(\mathbf{v}_f-\mathbf{v}_n\right)=\mathbf{0},  \\
 \mathrm{Momentum\, Eq.\, for\, the\, fluid:} & \qquad (1-\phi) \nabla p +\xi \left(\mathbf{v}_f-\mathbf{v}_n\right)=\mathbf{0}, 
\end{align}
\end{subequations}
where subscripts $n$ and $f$ denote network and fluid phases, $\phi$ is the volume fraction of the network phase, $\mathbf{\sigma}_n$ is the network stress, $p$ is the pressure 
that ensures the incompressibility constraint in Eq.~(\ref{eq:biot}$a$), $\xi$ is the friction constant per unit volume of the material  and $\mathbf{I}$ is the identity matrix. 
Using a general linear viscoelastic  (VE) constitutive equation (CE) for the network stress, $\mathbf{\sigma}_n$, leads to a system of \emph{Linear} initial/boundary value problems. The linearity of the governing equations has facilitated a number of analytical solutions in simple geometries (see \cite{detournay1993fundamentals} and Chapter 7 of \cite{cheng2016poroelasticity}). It has also led to the development of a reciprocal theorem and a boundary integral formulation of the equations of poroelasticity (see \cite{cheng1984boundary} and Chapter 8 of \cite{cheng2016poroelasticity}). 

However, Biot theory has one major limitation as it pertains to modeling filamentous biological networks: As seen in Eq.~(\ref{eq:biot}$c$), the theory neglects the fluid's viscous stresses and describes the fluid motion using Darcy law. Scaling analysis shows that --while they can be safely ignored for non-dilute and connected solid structures, such as rocks and soils-- the viscous forces cannot be neglected at very low volume fractions of the solid phase, specially for fibrous networks such as the cytoskeleton, and the extra cellular matrix. In these limits the fluid flows are more accurately described by Brinkman equation \citep{auriault2009domain, levy1983fluid, strychalski2016intracellular}.
The aim of this work is to develop the same set of tools and machinery used in Biot poroelasticity \citep{cheng2016poroelasticity}, linear elasticity and viscous flows \citep{kim2013microhydrodynamics, happel2012low}, for the more generalized equations of linear poroelasticity, where the viscous stresses are accounted for and fluid flows are modeled using Brinkman equation. We begin by deriving the general solution to the governing equations in spherical coordinate, using general linear VE-CEs for both the network and fluid phases and a constant permeability coefficient.  

As an example, we use the axisymmetric form of the general solution to study the dynamics of a rigid sphere moving in a poroelastic material, composed of a linear elastic network and a Newtonian fluid, under an external force.  We derive closed-form expressions for the time-dependent fluid and network displacement fields, and calculate the time-dependent mobility of the sphere, as a function of constitutive parameters of the fluid and the network. The motivation for studying this problem is to develop a formulation for characterizing poroelastic materials using single-particle and two-point particle-tracking microrheology (MR) under thermal (passive MR) and external (active MR) forces \citep{mason1995optical,squires2010fluid}. 
We show that the network compressibility, and the response of the fluid to that,  introduces a slow relaxation time-scale in addition to the typical relaxation time characterized as the ratio of fluid viscosity to network shear modulus.   Interestingly, this slow relaxation results in a non-monotonic displacement with time away from the probe, which has no analog in VE materials.  Altogether, our solution to the sphere problem provides a framework for differentiating between VE and poroelastic materials and to characterize fluid permeability in addition to VE properties of the fluid and the network from particle-tracking MR experiments.
%
%

%
\section{General solutions of equations of poroelasticity}
We begin by adding the contribution of the viscous shear stress in the momentum equation of the Newtonian fluid phase, which modifies Eq.~\eqref{eq:biot} to
\begin{subequations} \label{eq:PE}
\begin{align}
 \mathrm{Mass \,conservation:} & \qquad \nabla \cdot \left(\phi \mathbf{v}_n+(1-\phi)\mathbf{v}_f\right)=0  \label{eq:PEa} \\
 \mathrm{Momentum\, Eq.\, for\, the\, network:} & \qquad \nabla \cdot \left(\boldsymbol{\sigma}_n-\phi p \mathbf{I}\right)-\xi \left(\mathbf{v}_f-\mathbf{v}_n\right)=\mathbf{0},  \label{eq:PEb}\\
 \mathrm{Momentum\, Eq.\, for\, the\, fluid:} & \qquad \eta \nabla^2 \mathbf{v}_f+\xi \left(\mathbf{v}_f-\mathbf{v}_n\right)+(1-\phi) \nabla p =\mathbf{0}.  \label{eq:PEc}
\end{align}
\end{subequations}
We take the volume fraction of the network phase, $\phi$, to remain constant in space and time. Thus, the compressibility of the network leads to nonzero divergence of the fluid velocity field: $\nabla \cdot \mathbf{v}_f =(\phi/(1-\phi))\nabla \cdot \left(\mathbf{v}_n\right) \neq 0$. However, this apparent fluid compressibility is not associated with  changes in fluid density and arises from fluid sources and sinks in response to changes in the change in volume of the network phase. As such, the Newtonian fluid stress does not include the isotropic term that scales with $\eta_b \nabla \cdot \mathbf{v}_f$, where $\eta_b$ is the bulk viscosity. We use a general linear VE-CE to describe the traceless component of the fluid stress: 
\begin{equation}
{\boldsymbol{\sigma}}_f =  {\int}_0^t  G_f(t-t^\prime)  \left( \nabla {\mathbf{v}}_f (t^\prime)+ \nabla {\mathbf{v}}_f^{T} (t^\prime) \right) \mathrm{d} t^\prime, 
\label{eq:FCE}
\end{equation} where $G_f(t)$ is the fluid's shear modulus and superscript $T$ denotes transpose operation. 
Similarly, we model the network stress, $\boldsymbol{\sigma}_n$, using a general linear VE isotropic CE: 
\begin{eqnarray}
&&  {\boldsymbol{\sigma}}_n =  {\int}_0^t \left[ G(t-t^\prime)  \left( \nabla {\mathbf{v}}_n (t^\prime)+ \nabla {\mathbf{v}}_n^{T} (t^\prime) \right) +
\lambda(t-t^\prime)\left(\nabla \cdot \mathbf{v}_n(t^\prime)\right)\mathbf{I} \right] \mathrm{d} t^\prime, 
\label{eq:PECE}
\end{eqnarray} 
where, $G(t)$ and $\lambda(t)$ are the time-dependent first and second Lam\'e coefficients. For a linear elastic network these coefficients are related through the Poisson ratio, $\nu$: $\lambda=\frac{2G\nu}{1-2\nu}$. Taking Laplace transform of Eqs. \eqref{eq:FCE} and \eqref{eq:PECE} yields 
$\tilde{\boldsymbol{\sigma}_f}=\tilde{\eta}(s) \left(\nabla \tilde{\mathbf{v}}_f+\nabla \tilde{\mathbf{v}}^T_f\right)$ and
$\tilde{\boldsymbol{\sigma}}_n(s)=\tilde{G}(s)\left(\nabla \tilde{\mathbf{v}}_n+\nabla \tilde{\mathbf{v}}^T_n\right)+\tilde{\lambda}(s)\left(\nabla \cdot \tilde{\mathbf{v}}_n\right)$, where superscripts $\sim$ denote variables in Laplace space and $\tilde{\eta}(s)=\mathcal{L}[G_f(t)]$.

Substituting this expression in Eqs.~\eqref{eq:PE} in Laplace space gives
\begin{subequations} \label{eq2}
\begin{align}
&  \nabla \cdot \left( {\tilde{\mathbf{v}}}_n \phi +  {\tilde{\mathbf{v}}}_f (1- \phi)   \right) =0 , \label{eq2a} \\
 &  \tilde{\eta} (s){\nabla}^2 {\tilde{\mathbf{v}}}_f   + \xi (\tilde{\mathbf{v}}_n - \tilde{\mathbf{v}}_f  ) - (1- \phi )  \nabla \tilde{p} =\mathbf{0} , \label{eq2b}  \\
 & \tilde{G}(s)  {\nabla}^2  \tilde{\mathbf{v}}_n + \left( \tilde{\lambda}(s) +\tilde{G}(s)\right) \nabla (\nabla \cdot  \tilde{\mathbf{v}}_n ) - \xi ( \tilde{\mathbf{v}}_n - \tilde{\mathbf{v}}_f ) - \phi \nabla \tilde{p}=\mathbf{0}.  \label{eq2c} 
\end{align}
\end{subequations}
\\
Next, we introduce $q:= -\nabla \cdot {\mathbf{v}}_n$. Using Eq.~\eqref{eq2a}, we get 
$\nabla \cdot \mathbf{v}_f= \varepsilon  q$, where $\varepsilon = \frac{\phi}{\phi -1} $.
We, then, take divergence of Eqs.(\ref{eq2b}), and (\ref{eq2c}), and add them to get
\begin{equation}
{\nabla}^2 \Phi =0 ,\quad
\text{where}\quad \Phi := \left( \varepsilon \,\tilde{\eta} + (\tilde{\lambda} +2\tilde{G}) \right) q + \tilde{p} . \label{eq3} 
\end{equation} 
The general solution to Laplace equation in spherical coordinate is 
\begin{equation}
    \Phi(r, \theta,\varphi)=\sum_{\ell=0}^\infty \sum_{m=-\ell}^{m=\ell}\left[D_{\ell ,m}^{\pm} \begin{pmatrix}
r^{\ell} \\ r^{-\ell -1}
\end{pmatrix}\right] Y_{\ell ,m}(\theta,\varphi),\, \text{where} \quad Y_{\ell, m}=P_{\ell}^m (\cos \theta) \, {\mathrm{e}}^{i m \varphi}
\end{equation}
is the scalar spherical harmonic function of degree $\ell$ and order $m$, $P_{\ell}^m$ is the associated Legendre function of degree $\ell$ and order $m$, $\theta \in [0,\pi]$ is the polar angle and $\varphi \in [0,2\pi)$ is the azimuthal angle; $r^{\ell}$ and $r^{-\ell-1}$ are two linearly independent functions describing the variation of $\Phi$ in radial direction. For brevity, we have presented these two functions in the array format $()^{\pm}$, where $D^+$ ($D^-$) are the coefficients associated with the function in the first (second) row, to be determined through imposing boundary conditions (BCs). The first row contains the internal solutions (the functions are finite when $r\to0$ and are unbounded as $r\to \infty$), and the second row contains the external solutions (the functions decay to zero as $r\to \infty$ and are singular as $r\to \infty$). This notation is used throughout the paper. 

Eliminating the pressure term after applying the divergence operator to network and fluid phases produces a modified Helmholtz equation for $q$:
\begin{equation}
{\nabla}^2 q - {\alpha}_{\varepsilon}^2  q=0,
\quad \text{where} \quad {\alpha}_{\varepsilon}^2(s) =  \frac{\xi {(1- \varepsilon)}^2}{ \tilde{\lambda}(s) +2\tilde{G}(s) + {\varepsilon}^2  \, \tilde{\eta} }.\label{eq4}
\end{equation}
The general solution for $q$, in spherical coordinate can be expressed as
\begin{equation}
q(r , \theta,\varphi)=\sum_{\ell=0}^\infty \sum_{m=-\ell}^{m=\ell}\left[ E_{\ell ,m}^{\pm}
   \begin{pmatrix}
{\mathsf{i}}_{\ell } ({\alpha}_{\varepsilon} r) \\  {\mathsf{k}}_{\ell } ({\alpha}_{\varepsilon} r) 
\end{pmatrix}  \right] Y_{\ell ,m} (\theta , \varphi), 
\end{equation}
where ${\mathtt{i}}_{\ell } ({\alpha}_{\varepsilon} r)=  \sqrt{\frac{\pi}{2 {\alpha}_{\varepsilon} r}} I_{\ell +\frac{1}{2}} ({\alpha}_{\varepsilon} r) $,
${\mathtt{k}}_{\ell } ({\alpha}_{\varepsilon} r)=  \sqrt{\frac{2}{\pi {\alpha}_{\varepsilon} r}} K_{\ell +\frac{1}{2}} ({\alpha}_{\varepsilon} r)$, 
are modified spherical Bessel functions of first and second kind, respectively.
Taking the general solution of Laplace equation for $\Phi$, and modified Helmholtz equation for $q$, in spherical coordinates, we can express the pressure as:
\begin{equation}
\tilde{p} (r, \theta , \varphi)= \sum_{\ell=0}^\infty \sum_{m=-\ell}^{m=\ell}   \Bigg[ D_{\ell ,m}^{\pm} \begin{pmatrix}
r^{\ell} \\ r^{-\ell -1}
\end{pmatrix} - \Big( \varepsilon \tilde{\eta} + \tilde{\lambda} +2\tilde{G} \Big)  \begin{pmatrix}
{\mathsf{i}}_{\ell } ({\alpha}_{\varepsilon} r) \\  {\mathsf{k}}_{\ell } ({\alpha}_{\varepsilon} r) 
\end{pmatrix} E_{\ell ,m}^{\pm} \Bigg] Y_{\ell ,m} (\theta , \varphi). \label{eq5}
\end{equation}
\\
Next, we introduce ${\mathbf{v}}^{+} := \tilde{\eta} {\tilde{\mathbf{v}}}_f + \tilde{G} {\tilde{\mathbf{v}}}_n $, and ${\mathbf{v}}^{-} := {\tilde{\mathbf{v}}}_f  - {\tilde{\mathbf{v}}}_n$. After summation and subtraction of Eqs \eqref{eq2b} and \eqref{eq2c} and a few lines of algebra we arrive at the following equations for $\mathbf{v}^{+}$ and $\mathbf{v}^{-}$: 
\begin{subequations}
\begin{align}
& {\nabla}^2 {\mathbf{v}}^{+}  -  \nabla (\nabla \cdot {\mathbf{v}}^{+}) = \nabla \Phi, 
\label{eq7}\\
& {\nabla}^2  {\mathbf{v}}^{-} + \frac{{\gamma}_{\varepsilon}}{1- \varepsilon} \nabla (\nabla \cdot  {\mathbf{v}}^{-} ) - {\beta}^2 {\mathbf{v}}^{-} = \frac{1}{{\tilde{\eta}}_{\varepsilon}} \nabla \Phi, \label{eq8}
\end{align}
\end{subequations}
where ${ {\beta}^2 =  \xi (\frac{1}{\tilde{G}} + \frac{1}{\tilde{\eta}}) }$, $  \displaystyle{ \frac{1}{{\tilde{\eta}}_{\varepsilon}} = \frac{1}{1-\varepsilon} \left( \frac{1}{\tilde{\eta}} + \frac{\varepsilon}{\tilde{G}} \right) }$, ${\gamma}_{\varepsilon} ={  \frac{\tilde{\lambda} +\tilde{G}}{\tilde{G}}   + \frac{\tilde{\eta} }{1- \varepsilon} \left[  (\varepsilon + \frac{\tilde{\lambda} +2\tilde{G}}{\tilde{\eta} } ) (\frac{1}{\tilde{\eta}} + \frac{ \varepsilon}{\tilde{G}}) \right]   }$,
and ${     \frac{{\beta}^2 (1- \varepsilon)}{1- \varepsilon + {\gamma}_{\varepsilon}} = {\alpha}_{\varepsilon}^2   }$.
The general solutions for $ {\mathbf{v}}^{+} $ and $ {\mathbf{v}}^{-} $ are:
\be 
\begin{aligned}
& {\mathbf{v}}^{+} = \sum_{\ell ,m}     \begin{pmatrix}
r^{\ell} \\ r^{-\ell -1}
\end{pmatrix} \Bigg\{  A_{\ell ,m}^{\pm} \sqrt{\ell (\ell +1)} {\mathbf{C}}_{\ell ,m} 
+  B_{\ell ,m}^{\pm}      \begin{pmatrix}
(\ell +1) {\mathbf{P}}_{\ell +1,m} + \sqrt{(\ell +1)(\ell +2)} {\mathbf{B}}_{\ell +1,m} \\
- \ell {\mathbf{P}}_{\ell -1,m} + \sqrt{\ell (\ell -1)} {\mathbf{B}}_{\ell -1,m}
\end{pmatrix} \Bigg\}  \qquad\qquad\qquad  \\
& \qquad\qquad  + D_{\ell ,m}^{\pm}  \begin{pmatrix}
r^{\ell +1} \\ r^{-\ell } \end{pmatrix} \begin{pmatrix}
\frac{1}{2(2\ell +3)} \\ \frac{1}{2(2\ell -1)}
\end{pmatrix}
\begin{pmatrix} 
 \ell {\mathbf{P}}_{\ell ,m} + \frac{\ell +3 }{\ell +1} \sqrt{\ell (\ell +1)} {\mathbf{B}}_{\ell ,m} \\
(\ell +1) {\mathbf{P}}_{\ell ,m} + \frac{2- \ell}{ \ell} \sqrt{\ell (\ell +1)} {\mathbf{B}}_{\ell ,m} 
\end{pmatrix}      \\
 & \qquad\qquad  \displaystyle{   - \frac{\varepsilon \tilde{\eta} +\tilde{G}}{{\alpha}_{\varepsilon}^2} \,  E_{\ell ,m}^{\pm} \left[  \frac{\mathrm{d}}{\mathrm{d} r}   \begin{pmatrix}
{\mathsf{i}}_{\ell } ({\alpha}_{\varepsilon} r) \\  {\mathsf{k}}_{\ell } ({\alpha}_{\varepsilon} r) 
\end{pmatrix}  {\mathbf{P}}_{\ell ,m}+ \frac{1}{r} \begin{pmatrix}
{\mathsf{i}}_{\ell } ({\alpha}_{\varepsilon} r) \\  {\mathsf{k}}_{\ell } ({\alpha}_{\varepsilon} r) 
\end{pmatrix}  \sqrt{\ell(\ell +1)}  {\mathbf{B}}_{\ell ,m}  \right] ,  }                     
\end{aligned} \label{eq9}
\ee
\\
\be
\begin{aligned}
&  {\mathbf{v}}^{-} =    \sum_{\ell ,m} \Bigg\{      {A'}_{\ell ,m}^{\pm}    \begin{pmatrix}
{\mathsf{i}}_{\ell } (\beta r) \\  {\mathsf{k}}_{\ell } (\beta r) 
\end{pmatrix} \sqrt{\ell (\ell +1)} {\mathbf{C}}_{\ell ,m}  +  {B '}_{\ell ,m}^{\pm}  \Bigg[  \ell (\ell +1) {\mathbf{P}}_{\ell ,m} \frac{1}{\beta r}  \begin{pmatrix}
{\mathsf{i}}_{\ell } (\beta r) \\  {\mathsf{k}}_{\ell } (\beta r) 
\end{pmatrix} +    \sqrt{\ell (\ell +1)}  {\mathbf{B}}_{\ell ,m}  \Bigg(    \frac{\mathrm{d}}{\mathrm{d} (\beta r)}   \times        \\
& \qquad\qquad \begin{pmatrix}
{\mathsf{i}}_{\ell } (\beta r) \\  {\mathsf{k}}_{\ell } (\beta r) 
\end{pmatrix}  + \frac{1}{\beta r}  \begin{pmatrix}
{\mathsf{i}}_{\ell } (\beta r) \\  {\mathsf{k}}_{\ell } (\beta r) 
\end{pmatrix}  \Bigg)   \Bigg]    - \frac{1}{ {\tilde{\eta}}_{\varepsilon} {\beta}^2} {D}_{\ell ,m}^{\pm} \begin{pmatrix}
r^{\ell -1} \\ r^{-\ell -2}
\end{pmatrix}  \begin{pmatrix}
\ell {\mathbf{P}}_{\ell ,m} + \sqrt{\ell (\ell +1)} {\mathbf{B}}_{\ell ,m} \\
-(\ell +1) {\mathbf{P}}_{\ell ,m} + \sqrt{\ell (\ell +1)} {\mathbf{B}}_{\ell ,m}
\end{pmatrix}       \\
& \qquad\qquad     - \frac{ \varepsilon -1 }{ {\alpha}_{\varepsilon}^2 }  {E}_{\ell ,m}^{\pm}  \Bigg[ \frac{\mathrm{d}}{\mathrm{d} r}  \begin{pmatrix}
{\mathsf{i}}_{\ell } ( {\alpha}_{\varepsilon} r) \\  {\mathsf{k}}_{\ell } ( {\alpha}_{\varepsilon} r) 
\end{pmatrix} {\mathbf{P}}_{\ell ,m}  + \frac{1}{r}    \begin{pmatrix}
{\mathsf{i}}_{\ell } ( {\alpha}_{\varepsilon} r) \\  {\mathsf{k}}_{\ell } ( {\alpha}_{\varepsilon} r) 
\end{pmatrix} \sqrt{\ell (\ell +1)}  {\mathbf{B}}_{\ell ,m}         \Bigg]                 \Bigg\}     .   
\end{aligned} \label{eq10}
\ee
\begin{subequations}
\begin{align}
\mathrm{where}\qquad  {\mathbf{P}}_{\ell ,m} &= Y_{\ell ,m} (\theta ,\varphi) \hat{\mathbf{r}} , \label{eq11} \\
 {\mathbf{B}}_{\ell ,m} &=\frac{1}{\sqrt{\ell (\ell +1)}} \left[ \frac{\partial }{\partial \theta} \hat{\boldsymbol{\theta}} + \frac{1}{\sin \theta}   \frac{\partial }{\partial \varphi} \hat{\boldsymbol{\varphi}}  \right] Y_{\ell,m} (\theta , \varphi) ,  \label{eq12}  \\
{\mathbf{C}}_{\ell ,m} &=  \frac{1}{\sqrt{\ell (\ell +1)}} \left[  \frac{1}{\sin \theta}  \frac{\partial }{\partial \varphi}   \hat{\boldsymbol{\theta}}  -   \frac{\partial }{\partial \theta}   \hat{\boldsymbol{\varphi}}  \right] Y_{\ell,m} (\theta , \varphi)   \label{eq13}
\end{align}
\end{subequations}
are scaloidal-poloidal-toroidal representation of mutually orthogonal vector spherical harmonics that are commonly used to express the solutions to vector Laplace and Helmholtz equations in spherical coordinates \cite{morse1953methods}.  $A_{\ell ,m}^{\pm}$, $B_{\ell ,m}^{\pm}$, ${A'}_{\ell ,m}^{\pm}$, ${B'}_{\ell ,m}^{\pm}$, $D_{\ell ,m}^{\pm}$, and $E_{\ell ,m}^{\pm}$ are 12 constants to be determined from BCs; see Appendix~\ref{appA} for solution details. Substituting these expressions in relations  ${\tilde{\mathbf{v}}}_f = \frac{1}{\tilde{\eta} + \tilde{G}} ({\mathbf{v}}^{+} + \tilde{G} {\mathbf{v}}^{-} )$ and ${\tilde{\mathbf{v}}}_n =           \frac{1}{\tilde{\eta} + \tilde{G}} ({\mathbf{v}}^{+} - \tilde{\eta} {\mathbf{v}}^{-} )$ gives the fluid and network velocity (or displacement) fields in S-space, which we do not produce here for brevity.

\subsection{Axisymmetric Solutions}
The general solutions to $\mathbf{v}_f$ and $\mathbf{v}_n$ can be used to obtain analytical solutions to a variety of problems involving spherical geometries. Several applications of Lamb's general solution for Stokes flow may be found in \cite{happel2012low} and \cite{kim2013microhydrodynamics}; the analytical solutions to poroelasticity equation using Biot's general solution are surveyed in \cite{detournay1993fundamentals} and \cite{cheng2016poroelasticity}. The majority of these solutions involve axisymmetric geometries i.e. $\partial (v_\theta,v_r) /\partial \phi =0$, ${v}_\phi=0$. This is achieved when the summation index $m=0$ in 
Eqs.~(\ref{eq9}) and (\ref{eq10}), and ${A}_{\ell m}^{\pm}={A'}_{\ell m}^{\pm}=0$. 
The axisymmetric forms of fluid and network velocities simplify to:
\begin{subequations}
\begin{align}
& \tilde{v}_{r,f} = \frac{1}{\tilde{\eta} +\tilde{G}} \Bigg[  \sum_{\ell =0 }^{\infty} \Bigg\{ \begin{pmatrix}
 {B}_{\ell -1}^{+} \\ {B}_{\ell +1}^{-} \end{pmatrix}  \begin{pmatrix}
 \ell r^{\ell -1} \\ -(\ell +1) r^{-\ell -2} 
 \end{pmatrix} + {B'}_{\ell}^{\pm} \frac{\ell (\ell +1) \tilde{G}}{\beta r}  \begin{pmatrix}
{\mathsf{i}}_{\ell } (\beta r) \\  {\mathsf{k}}_{\ell } (\beta r) 
\end{pmatrix}     \label{eq23a}  \\
& \qquad\qquad\qquad  + D_{\ell }^{\pm} \begin{pmatrix}
r^{\ell -1} \\ r^{-\ell -2}
\end{pmatrix} \begin{pmatrix} 
\frac{\ell}{2(2\ell +3)} r^2 - \frac{\tilde{G}}{{\tilde{\eta}}_{\varepsilon}  {\beta}^2} \ell \\
\frac{\ell +1}{2(2\ell -1)} r^2 + \frac{\tilde{G}}{ {\tilde{\eta}}_{\varepsilon}  {\beta}^2} (\ell +1)   
\end{pmatrix}  - \frac{\varepsilon (\tilde{\eta} +\tilde{G})}{{\alpha}_{\varepsilon}^2} {E}_{\ell}^{\pm} \frac{\mathrm{d}}{\mathrm{d}r}  
 \begin{pmatrix}
{\mathsf{i}}_{\ell } ( {\alpha}_{\varepsilon} r) \\  {\mathsf{k}}_{\ell } ( {\alpha}_{\varepsilon} r) 
\end{pmatrix} \Bigg\} P_{\ell} (\cos\theta) \Bigg] , \nn  \\
& \tilde{v}_{\theta ,f} =  \frac{1}{\tilde{\eta} +\tilde{G}} \Bigg[  \sum_{\ell =1 }^{\infty} \Bigg\{ \begin{pmatrix}
 {B}_{\ell -1}^{+} \\ {B}_{\ell +1}^{-} \end{pmatrix}  \begin{pmatrix}
  r^{\ell -1} \\ r^{-\ell -2} 
 \end{pmatrix} + {B'}_{\ell}^{\pm} \frac{\tilde{G}}{\beta } \Big[ \frac{\mathrm{d}}{\mathrm{d}r} \begin{pmatrix}
{\mathsf{i}}_{\ell } (\beta r) \\  {\mathsf{k}}_{\ell } (\beta r) 
\end{pmatrix} + \frac{1}{r}  \begin{pmatrix}
{\mathsf{i}}_{\ell } (\beta r) \\  {\mathsf{k}}_{\ell } (\beta r) 
\end{pmatrix}  \Big]   \label{eq23b}  \\
& \qquad\qquad\qquad   + D_{\ell }^{\pm} \begin{pmatrix}
r^{\ell -1} \\ r^{-\ell -2}
\end{pmatrix} \begin{pmatrix}
\frac{\ell +3}{2(\ell +1)(2\ell +3)} r^2 - \frac{\tilde{G}}{{\tilde{\eta}}_{\varepsilon} {\beta}^2}  \\
\frac{2- \ell}{2 \ell (2\ell -1)} r^2 - \frac{\tilde{G}}{{\tilde{\eta}}_{\varepsilon} {\beta}^2} 
\end{pmatrix}       - \frac{\varepsilon (\tilde{\eta} +\tilde{G})}{{\alpha}_{\varepsilon}^2} {E}_{\ell}^{\pm} \frac{1}{r} 
 \begin{pmatrix}
{\mathsf{i}}_{\ell } ( {\alpha}_{\varepsilon} r) \\  {\mathsf{k}}_{\ell } ( {\alpha}_{\varepsilon} r) 
\end{pmatrix} \Bigg\} \frac{\mathrm{d}}{\mathrm{d}\theta} P_{\ell} (\cos\theta) \Bigg] , \nn 
\end{align}
\end{subequations}

\begin{subequations}
\begin{align}
& \tilde{v}_{r,n} = \frac{1}{\tilde{\eta} +\tilde{G}} \Bigg[  \sum_{\ell =0 }^{\infty} \Bigg\{ \begin{pmatrix}
 {B}_{\ell -1}^{+} \\ {B}_{\ell +1}^{-} \end{pmatrix}  \begin{pmatrix}
 \ell r^{\ell -1} \\ -(\ell +1) r^{-\ell -2} 
 \end{pmatrix} - {B'}_{\ell}^{\pm} \frac{\ell (\ell +1) \tilde{\eta}}{\beta r}  \begin{pmatrix}
{\mathsf{i}}_{\ell } (\beta r) \\  {\mathsf{k}}_{\ell } (\beta r) 
\end{pmatrix}     \label{eq24a}  \\
& \qquad\qquad\qquad  + D_{\ell }^{\pm} \begin{pmatrix}
r^{\ell -1} \\ r^{-\ell -2}
\end{pmatrix} \begin{pmatrix} 
\frac{\ell}{2(2\ell +3)} r^2 + \frac{\tilde{\eta}}{{\tilde{\eta}}_{\varepsilon} {\beta}^2} \ell \\
\frac{\ell +1}{2(2\ell -1)} r^2 - \frac{\tilde{\eta}}{{\tilde{\eta}}_{\varepsilon} {\beta}^2} (\ell +1)   
\end{pmatrix}  -  \frac{  \tilde{G} + \tilde{\eta} }{{\alpha}_{\varepsilon}^2} {E}_{\ell}^{\pm} \frac{\mathrm{d}}{\mathrm{d}r}  
 \begin{pmatrix}
{\mathsf{i}}_{\ell } ( {\alpha}_{\varepsilon} r) \\  {\mathsf{k}}_{\ell } ( {\alpha}_{\varepsilon} r) 
\end{pmatrix} \Bigg\} P_{\ell} (\cos\theta) \Bigg] , \nn  \\ 
& \tilde{v}_{\theta ,n} =  \frac{1}{\tilde{\eta} +\tilde{G}} \Bigg[  \sum_{\ell =1 }^{\infty} \Bigg\{ \begin{pmatrix}
 {B}_{\ell -1}^{+} \\ {B}_{\ell +1}^{-} \end{pmatrix}  \begin{pmatrix}
  r^{\ell -1} \\ r^{-\ell -2} 
 \end{pmatrix} - {B'}_{\ell}^{\pm} \frac{\tilde{\eta}}{\beta } \Big[ \frac{\mathrm{d}}{\mathrm{d}r} \begin{pmatrix}
{\mathsf{i}}_{\ell } (\beta r) \\  {\mathsf{k}}_{\ell } (\beta r) 
\end{pmatrix} + \frac{1}{r}  \begin{pmatrix}
{\mathsf{i}}_{\ell } (\beta r) \\  {\mathsf{k}}_{\ell } (\beta r) 
\end{pmatrix}  \Big]   \label{eq24b}   \\
& \qquad\qquad\qquad   + D_{\ell }^{\pm} \begin{pmatrix}
r^{\ell -1} \\ r^{-\ell -2}
\end{pmatrix} \begin{pmatrix}
\frac{\ell +3}{2(\ell +1)(2\ell +3)} r^2 + \frac{\tilde{\eta}}{{\tilde{\eta}}_{\varepsilon} {\beta}^2}  \\
\frac{2- \ell}{2 \ell (2\ell -1)} r^2 + \frac{\tilde{\eta}}{{\tilde{\eta}}_{\varepsilon} {\beta}^2} 
\end{pmatrix}       -  \frac{  \tilde{G} + \tilde{\eta} }{{\alpha}_{\varepsilon}^2} {E}_{\ell}^{\pm} \frac{1}{r} 
 \begin{pmatrix}
{\mathsf{i}}_{\ell } ( {\alpha}_{\varepsilon} r) \\  {\mathsf{k}}_{\ell } ( {\alpha}_{\varepsilon} r) 
\end{pmatrix} \Bigg\} \frac{\mathrm{d}}{\mathrm{d}\theta} P_{\ell} (\cos\theta) \Bigg] .  \nn  
\end{align}
\end{subequations}
We note that fluid velocity dependency on the network compressibility (through ${\alpha}_{\varepsilon}$) is proportional to $\phi \sim \epsilon \ll 1$ and, thus, negligible in most physiologically relevant conditions. 

\section{A rigid sphere moving in a poroelastic medium }
We use the axisymmetric solutions to study the dynamics of a spherical bead moving within a poroelastic medium. The motivation for studying this problem is to develop a mathematical framework that allows for measuring the 
fluid permeability of cellular networks by single and two-point particle tracking MR. 

The current framework for analyzing microrheological data is the 
generalized Stokes-Einstein relation (GSER), which predicates on the applicability of Stokes relation between external force and velocity in Laplace (or Fourier) space:
$\tilde{F}(s)=\tilde{R}(s) \tilde{U}(s)$, where $\tilde{R}(s)=6\pi \tilde{G}(s) a$ is the so called response function, or resistance, for a linear VE fluid. 
While this relation holds for linear VE materials, it is not generally applicable to PE materials. 
The domains of applicability of GSER in a PE material was  studied by \cite{levine2001response}, who 
approximated the solution to a sphere moving in poroelastic medium using the fundamental solutions to a point-force in a poroelastic material. Instead of directly imposing the BCs, the authors introduced a wave vector cutoff of $k_\text{max}=\pi/a$ to the fundamental solutions in Fourier domain. Here, we provide an exact solution of the equations by directly imposing the BCs on the sphere surface. Moreover, we discuss the differences between the dynamics of the sphere moving in VE and PE materials in the context of single-particle and two-point MR. 

Consider a rigid sphere moving with time-dependent velocity $U(t) \hat{\mathbf{z}}$ in an unbounded poroelastic medium. We assume that the size of the sphere is larger than the typical mesh size of the background network, and, thus, assume no-slip BCs for both the network and fluid velocity fields:
\begin{align}
&\text{at}  \, r=a: & \tilde{v}_{r,f}=\tilde{v}_{r,n}=\tilde{U}(s)\cos \theta,&  &\tilde{v}_{\theta,f}=\tilde{v}_{\theta,n} = -\tilde{U}(s)\sin \theta. &
\label{eq:no-slip}
\end{align}
Since the velocity field must decay to zero at infinity, and that functions corresponding to internal solutions are all unbounded as $r\to \infty$, the coefficients corresponding to internal solutions (all coefficients with $+$ superscript) are identically zero. 
Applying no-slip BCs, Eqs.~\eqref{eq:no-slip}, to the axisymmetric solutions for the fluid and network,  we get:
\begin{subequations} \label{eq17ab}
\begin{align}
 {\tilde{v}}_{r,f}&=  \frac{a}{{\beta}^2 r^3} \frac{\tilde{U}(s)}{{\Delta}_{\varepsilon}}  \Big[  g_1 + (1-\varepsilon)   {\tau}_{\varepsilon} (3 r^2 - a^2) {\beta}^2 g_2  -3 \frac{{\alpha}_{\varepsilon}^2}{{\beta}^2} (1-\varepsilon) \, e^{-\beta (r-a)} (1+ \beta r)  \,-\varepsilon g_3   \label{eq17a} \\
& \qquad\qquad + 3\varepsilon (1+\tau) e^{- {\alpha}_{\varepsilon} (r-a)} (2+ 2{\alpha}_{\varepsilon} r + {\alpha}_{\varepsilon}^2 r^2)  \Big] \cos\theta ,  \nn \\
   {\tilde{v}}_{\theta ,f} &=   \frac{a}{2{\beta}^2 r^3} \frac{\tilde{U}(s)}{{\Delta}_{\varepsilon}}  \Big[  g_1 - (1-\varepsilon)   {\tau}_{\varepsilon} (3 r^2 + a^2) {\beta}^2 g_2   -3 \frac{{\alpha}_{\varepsilon}^2}{{\beta}^2}  (1-\varepsilon) \, e^{-\beta (r-a)} (1+ \beta r + {\beta}^2 r^2) \,-\varepsilon g_3   \label{eq17b} \\
    &  \qquad\qquad  + 6\varepsilon (1+\tau) e^{- {\alpha}_{\varepsilon} (r-a)} (1+ {\alpha}_{\varepsilon} r )  \Big] \sin\theta , \nn \\
  {\tilde{v}}_{r,n} &=  \frac{a}{{\beta}^2 r^3} \frac{\tilde{U}(s)}{{\Delta}_{\varepsilon}}  \Big[  -g_4 + (1-\varepsilon)   {\tau}_{\varepsilon} (3 r^2 - a^2) {\beta}^2 g_2  +3 \frac{{\alpha}_{\varepsilon}^2}{{\beta}^2} \tau (1-\varepsilon) \, e^{-\beta (r-a)} (1+ \beta r)  \,+\varepsilon \, \tau g_1   \label{eq17c} \\
 & \qquad\qquad + 3 (1+\tau) e^{- {\alpha}_{\varepsilon} (r-a)} (2+ 2{\alpha}_{\varepsilon} r + {\alpha}_{\varepsilon}^2 r^2)  \Big] \cos\theta , \nn   \\
    {\tilde{v}}_{\theta ,n} &=   \frac{a}{2{\beta}^2 r^3} \frac{\tilde{U}(s)}{{\Delta}_{\varepsilon}}  \Big[  -g_4 - (1-\varepsilon)   {\tau}_{\varepsilon} (3 r^2 + a^2) {\beta}^2 g_2  +3 \frac{{\alpha}_{\varepsilon}^2}{{\beta}^2} \tau (1-\varepsilon) \, e^{-\beta (r-a)} (1+ \beta r + {\beta}^2 r^2)  \,+\varepsilon \, \tau g_1  \label{eq17d}   \\
 &\qquad\qquad + 6 (1+\tau) e^{- {\alpha}_{\varepsilon} (r-a)} (1+ {\alpha}_{\varepsilon} r )  \Big] \sin\theta ,  \nn
 \end{align}
 \end{subequations}
   where $\tau (s) = \frac{\tilde{\eta}(s)}{\tilde{G} (s)}$, $\tilde{\eta}_{\varepsilon} = \frac{(1-\varepsilon) \tilde{\eta}}{1+ \varepsilon \tau(s)}$, ${\tau}_{\varepsilon} (s)= \frac{\tilde{\eta}_{\varepsilon}}{\tilde{G}(s)}$. 
   Defining inverse of permeability as ${\beta}_{\circ}^2 = \frac{\xi}{\tilde{\eta}}$, we have
    ${\beta}^2 =  {\beta}_{\circ}^{2} \big(1+\tau (s) \big)$, ${\alpha}_{\varepsilon}^2 =  {\beta}_{\circ}^{2}  \frac{\tau (s) {(1-\varepsilon)}^2}{\frac{2(1-\nu)}{1-2\nu}  + {\varepsilon}^2 \tau (s)} $. The remaining coefficients that appear in Eqs.~\eqref{eq17ab}   are given by the following expressions:
\begin{subnumcases}{}
 {\Delta}_{\varepsilon} = 2 {\tau}_{\varepsilon} (1-\varepsilon) (1+ a {\alpha}_{\varepsilon}) + \frac{{\alpha}_{\varepsilon}^2}{{\beta}^2} \Big( 1+ \varepsilon \tau +  {\tau}_{\varepsilon}  (1-\varepsilon) (1+ a\beta + a^2 {\beta}^2)   \Big)    \label{eq19a} \\
g_1 =  \frac{{\alpha}_{\varepsilon}^2}{{\beta}^2} (3+3a \beta + a^2 {\beta}^2)    ,  \qquad   \qquad\qquad \qquad g_2 =   1+ a {\alpha}_{\varepsilon} + \frac{1}{2} \frac{{\alpha}_{\varepsilon}^2}{{\beta}^2} (1+ a\beta + a^2 {\beta}^2)     ,         \label{eq19b}     \\ 
g_3 =  6(1+\tau) (1+ a {\alpha}_{\varepsilon}) + 2 \tau a^2 {\alpha}_{\varepsilon}^2 + 3  \frac{{\alpha}_{\varepsilon}^2}{{\beta}^2}  (1+ a\beta + a^2 {\beta}^2)        ,  \label{eq19c}  \\
g_4 = 6(1+\tau) (1+ a {\alpha}_{\varepsilon}) + 2 a^2 {\alpha}_{\varepsilon}^2 + 3 \tau  \frac{{\alpha}_{\varepsilon}^2}{{\beta}^2}  (1+ a\beta + a^2 {\beta}^2)             . \label{eq19d} 
\end{subnumcases}

Substituting for coefficients in Eq. \eqref{eq5},  the pressure given by: 

\begin{eqnarray}
 \tilde{p} 
 =   \frac{3 a (\tilde{\eta} +\tilde{G})}{2 r^2} \frac{\tilde{U}(s)}{{\Delta}_{\varepsilon}} \Big[  2 {\tau}_{\varepsilon} (1-\varepsilon) (1+ a {\alpha}_{\varepsilon}) + {\tau}_{\varepsilon} \frac{{\alpha}_{\varepsilon}^2}{{\beta}^2} (1- \varepsilon) (1+a\beta + a^2 {\beta}^2) 
-2 \frac{{\alpha}_{\varepsilon}^2}{{\beta}^2}(\varepsilon \tau + \frac{\tilde{\lambda}}{\tilde{G}} +2 ) e^{-{\alpha}_{\varepsilon} (r-a)} (1+ {\alpha}_{\varepsilon} r)    \Big] \cos\theta . \nn \\
\end{eqnarray}

The fluid and network stress components, including the traceless and isotropic  parts, are: (see Appendix~\ref{appB} for an explicit expression of the network and fluid stresses.)
\begin{subequations}
\begin{align}
&\tilde{\sigma}_{rr}^{\mathrm{f}} = -\tilde{p} (1-\phi)+ 2 \tilde{\eta} \frac{\partial \tilde{v}_{r,\mathrm{f}}}{\partial r} , 
\qquad \qquad  \tilde{\sigma}_{rr}^{\mathrm{n}} 
 = (\tilde{\lambda} + 2\tilde{G})  \frac{\partial \tilde{v}_{n,r} }{\partial r} +  2 \tilde{\lambda}  \frac{\tilde{v}_{n,r} + \tilde{v}_{n, \theta} \cot\theta}{r}- \phi \tilde{p}  ,   \\
& \tilde{\sigma}_{r\theta}^{\mathrm{n}} 
= \tilde{G} \left( \frac{1}{r} \frac{\partial \tilde{v}_{n,r}}{\partial\theta} + \frac{\partial \tilde{v}_{n, \theta}}{\partial r} - \frac{\tilde{v}_{n, \theta}}{r}  \right), \qquad 
\tilde{\sigma}_{r\theta}^{\mathrm{f}} = \tilde{\eta} \left( \frac{1}{r} \frac{\partial \tilde{v}_{r,\mathrm{f}}}{\partial \theta} + \frac{\partial \tilde{v}_{\theta , \mathrm{f}}}{\partial r} - \frac{\tilde{v}_{\theta , \mathrm{f}}}{r}\right) .  \label{eq22}
\end{align}
\end{subequations}

Evaluating these stresses at $r=a$ and integrating over the sphere surface gives the following relation for the total force from the fluid and network phase: 
\begin{align}
& \tilde{\mathbf{F}}(s) 
= \int_{0}^{2\pi} \int_{0}^{\pi} \Big( ( \tilde{\sigma}_{rr}^{\mathrm{f}} + \tilde{\sigma}_{rr}^{\mathrm{n}}  ) \, \hat{\mathbf{r}} + (  \tilde{\sigma}_{r\theta}^{\mathrm{f}} +  \tilde{\sigma}_{r\theta}^{\mathrm{n}} ) \, \hat{\boldsymbol{\theta}} \Big) a^2 \, \sin\theta \, \mathrm{d}\theta \, \mathrm{d}\varphi    
=\tilde{R}(s)\tilde{U}(s)\hat{\mathbf{z}},  \label{eq23}  \\ 
&\tilde{R}(s)=  6\pi \tilde{G}(s) a \frac{\left(1+\tau\right)\tau}{{\Delta}_{\varepsilon}} \left[  \frac{\tilde{\eta}_{\varepsilon}}{\tilde\eta} (1-\varepsilon) \Big(     2(1+a {\alpha}_{\varepsilon} ) + \frac{{\alpha}_{\varepsilon}^2}{{\beta}^2} (1+ a\beta + a^2 {\beta}^2)  \Big) + \frac{2}{3}\varepsilon \frac{{\alpha}_{\varepsilon}^2}{{\beta}^2} (1+ a {\alpha}_{\varepsilon})  \right], \nn
\end{align}
where $\tilde{R}(s)$ is response function (or the hydrodynamic resistance) we set out to find. Using fluctuation dissipation theorem the measurable 
MSD can be related to the response function using 
$ \langle \Delta \tilde{\mathbf{r}}^2 \rangle =6k_{B}T \tilde{M}(s)/s^2$ \citep{squires2010fluid}, where $\langle \,\rangle$ denote ensemble averages, $\Delta \tilde{\mathbf{r}}^2$ is the MSD of the spherical probe, $\tilde{M}(s)=\tilde{R}^{-1}(s)$ is the sphere's isotropic mobility  in Laplace space, and $k_B T$ is Boltzmann thermal energy. This relation is used to compute $\tilde{R}(s)$, which in practice can be used  to compute $\tilde{G}(s)$, $\tilde{\eta}(s)$, $\nu$ and $\beta_\circ$ in single-particle tracking MR.

In two-point MR the positional cross correlation of two fluctuating probes are used to measure the properties of the medium  separating them. We can relate these movements to the medium's mechanical properties by applying  FDT, which yields: $ \mathcal{L}\langle \Delta \mathbf{x}_1(0)\Delta \mathbf{x}_2(t)\rangle=6 k_B T \tilde{\mathbf{M}}_{12}(s)/s^2$ \citep{squires2010fluid}, where $\tilde{\mathbf{M}}_{12}(s)$ is the pair mobility tensor that computes the velocity of particle $2$, due to a force on particle $1$: $\tilde{\mathbf{U}}_2(\mathbf{r},s)=\tilde{\mathbf{M}}_{12}(\mathbf{r},s)\cdot \tilde{\mathbf{F}}_1(s)$, where $\mathbf{r}$ is the pair separation vector. For VE materials, and when the particles are separated by a large distance ($r/a\gg 1$), the motion of particle 2 becomes independent of its size and asymptotes to the velocity of the surrounding VE fluid: $\tilde{\mathbf{U}}_2(\mathbf{r},s)\approx \tilde{\mathbf{v}}(\mathbf{r},s)$.

We use a similar approximation here and assume that particle 2 moves with the network velocity at its center: $\mathbf{U}_2(\mathbf{r})\approx \mathbf{v}_n(\mathbf{r})$ when $r/a\gg 1$. 
The pair mobility tensor can be expressed in the general form of $\tilde{\mathbf{M}}_{12}=\tilde{{M}}_{12}^\parallel \hat{\mathbf{r}}\hat{\mathbf{r}}+\tilde{{M}}_{12}^\perp \left(\mathbf{I}-\hat{\mathbf{r}}\hat{\mathbf{r}}\right)$, where $\parallel$ and $\perp$ denote parallel and perpendicular to the direction of applied force and $\mathbf{I}$ is the identity matrix, or Kronecker delta tensor.
Using $\tilde{U}(s)=\tilde{M}(s)\tilde{F}(s)$ in Eqs. \eqref{eq17ab} and using the above expression for mobility tensor, we arrive at the following expressions
\begin{align}
    & \tilde{M}_{12}^\parallel=\tilde{M}(s)\frac{a}{\beta^2 r^3 \Delta_\epsilon}[\dots]_{r,n}, & &
    \tilde{M}_{12}^\perp=\tilde{M}(s)\frac{a}{2 \beta^2 r^3 \Delta_\epsilon}[\dots]_{\theta,n},&
\end{align}
where $[\dots]_{r,n}$ and $[\dots]_{\theta,n}$ are the terms between brackets in Eqs.~\eqref{eq17c} and \eqref{eq17d}, respectively.
These expressions provide the mathematical framework for analyzing two-point MR results. 

\subsection{Results}
In this section we consider the example of a spherical probe moving under constant force ${\mathrm{F}}_{\circ}$, as an analog to active MR experiment, within a PE material composed of a linear elastic network, with shear modules $G$ and Poisson ratio $\nu$, and a Newtonian fluid of viscosity $\eta$, making $\tau(s)=s\eta/G$. The probe's velocity and the induced displacements in the network and fluid phases can be computed in Laplace space using Eqs.~(\ref{eq23}) and \eqref{eq17ab}, respectively. We, then, use Fourier-Euler summation \citep{abate1992fourier} to numerically invert the results from S- to time-space. 
The limiting values of all these quantities at $t=0$ and $t\to \infty$ can be computed analytically using the following limits in S-space: $f(t=0^+) =\lim_{s\to \infty} s\tilde{f}(s)$  and $f(t\to \infty) =\lim_{s\to 0} s\tilde{f}(s)$ (see Appendix~\ref{appC}).   
\begin{figure}[!t]
    \centering
     \subfigure[]{\includegraphics[width=0.45\textwidth , height=5.7cm ]{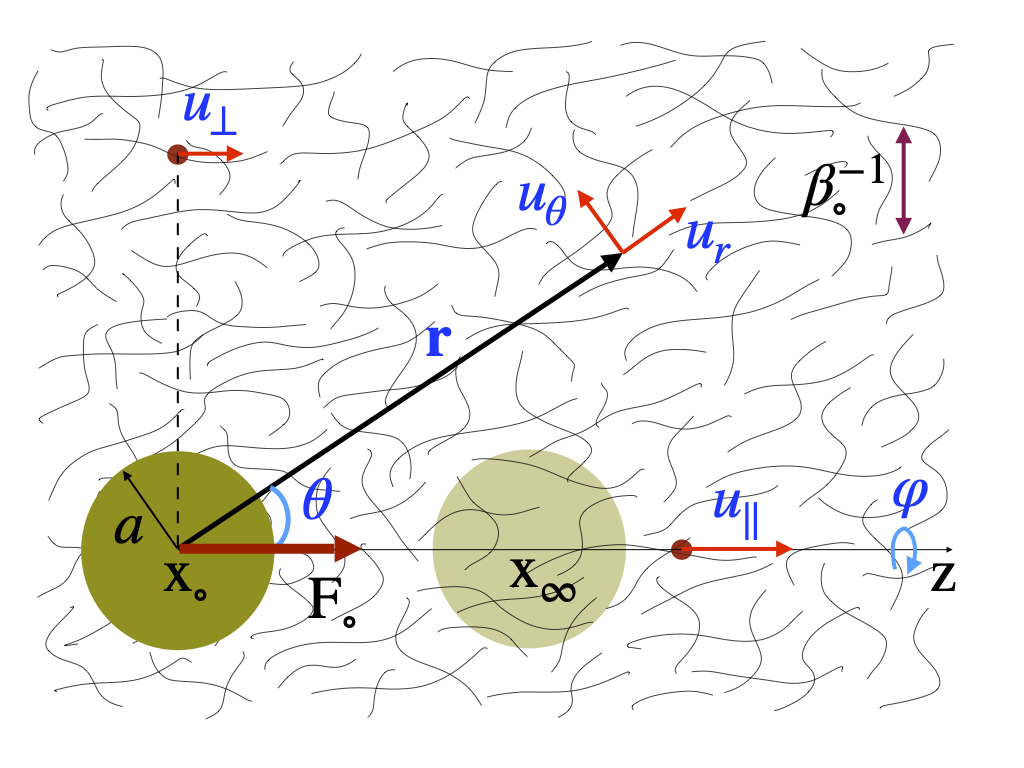}} \label{fig1a} \quad
    \subfigure[]{\includegraphics[width=0.49\textwidth , height=5.2cm ]{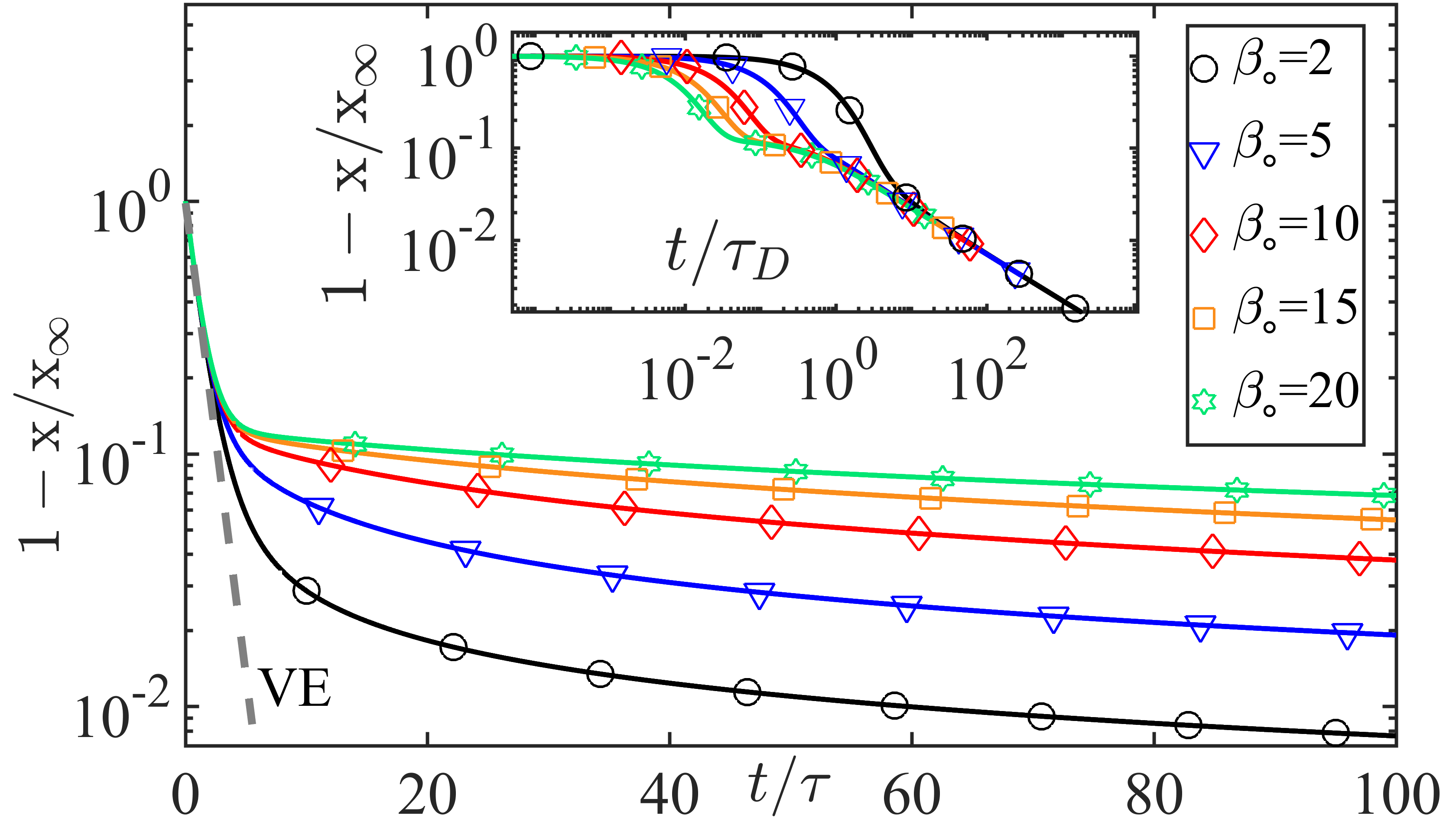}} \label{fig1b}
    \subfigure[]{\includegraphics[width=0.49\textwidth , height=5.2cm]{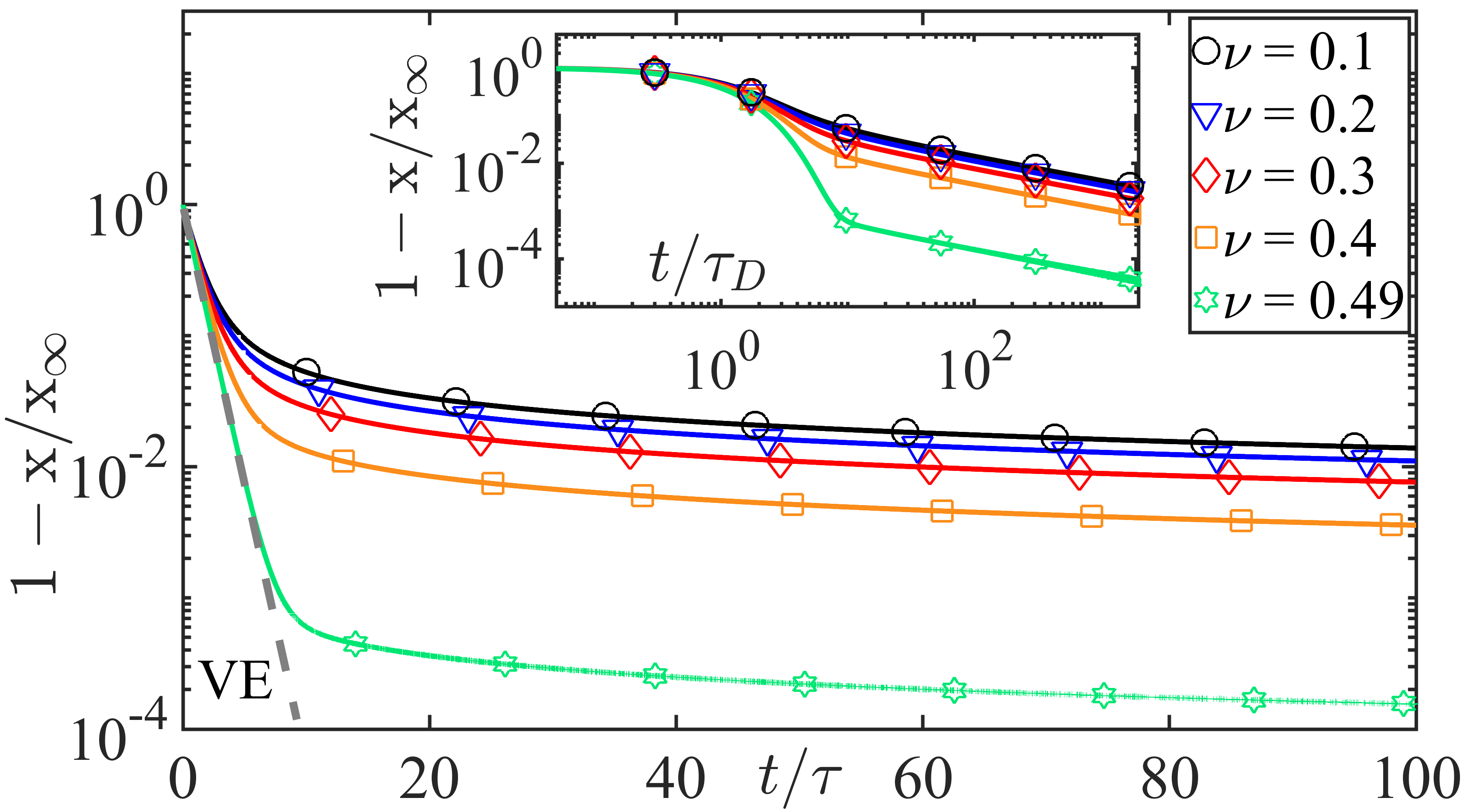}} \label{fig1c} %
    \subfigure[]{\includegraphics[width=0.49\textwidth , height=5.2cm]{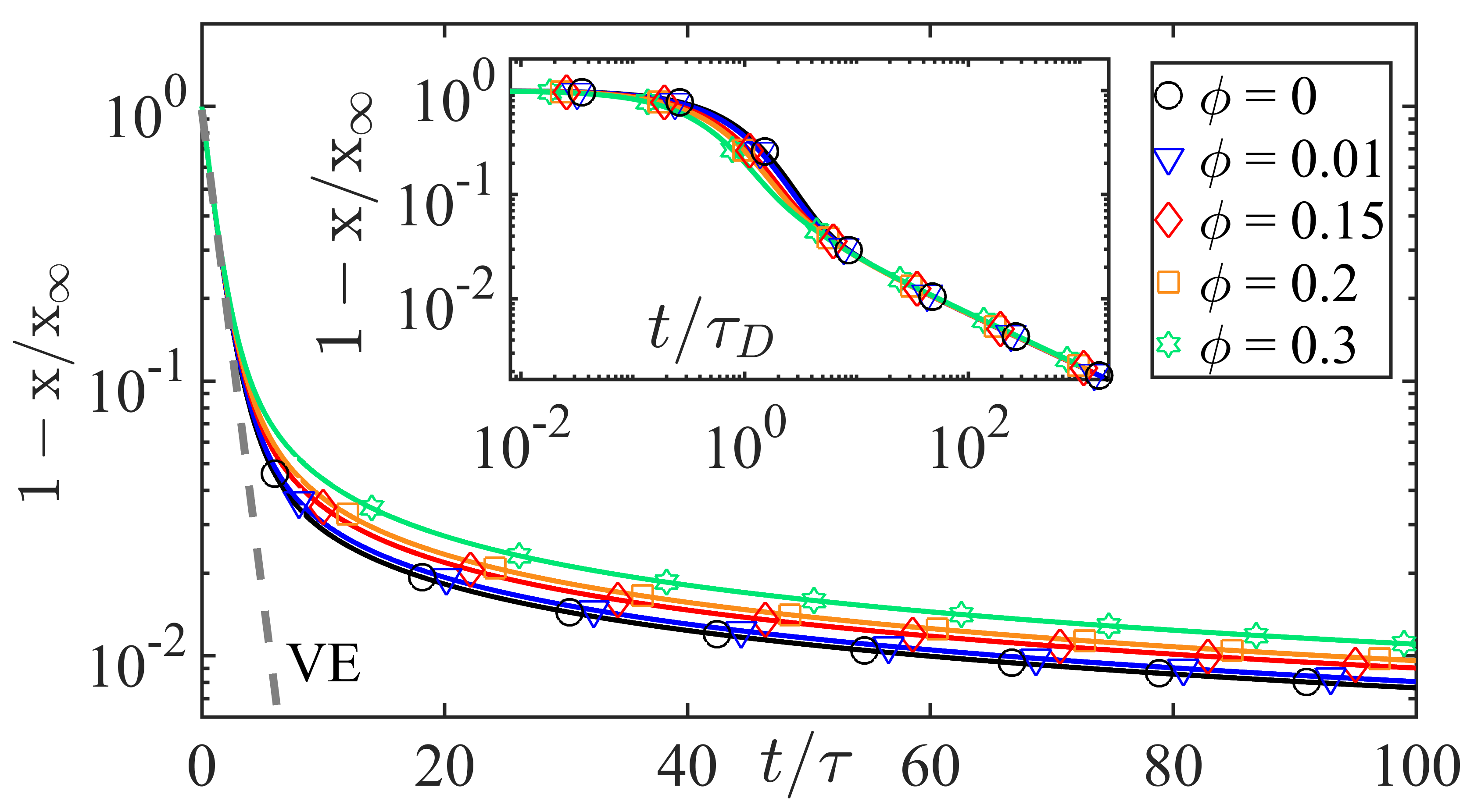}} \label{fig1d}
    \caption{  (a) A Schematic representation of a rigid spherical probe moving under a constant force in a poroelastic medium. The sphere moves from ${\mathrm{x}}_{\circ}$ at early times to ${\mathrm{x}}_{\infty}$ at long times.
    (b-d): The sphere relaxation $1-x(t)/x_\infty$, vs time for (b) different values of permeability, when $\phi \to 0$, and $\nu =0.3$, (c) different values of Poisson ratio, when $\phi \to 0$, ${\beta}_{\circ} =2$, and (d) different values of network volume fraction, when $\nu =0.3$, and ${\beta}_{\circ} =2$. The inset figures 
    represent the results of their associated main figure, as a function of rescaled time $t/\tau_D$, where $\tau_D=a^2/D_q$ is the diffusion timescale of compressibility field for a distance $a$, where $D_q=\tau^{-1}\alpha_\epsilon^{-2}$ is the diffusion coefficient of the network compressibility. 
    }
    \label{fig1}
\end{figure}
Figure~\ref{fig1}(b) shows the variations of 
$1-x/x_\infty$ vs $t/\tau$ for different values of $\beta_0>1$, when $\phi \to 0$ and $\nu=0.3$. Here, $x$ is the probe's position and ${x}_{\infty}=\frac{{F_{\circ}}}{6\pi G a} \frac{5-6\nu}{4(1-\nu)}$ is the long time position of the probe and $\tau=\eta/G$ is the 
time-scale of shear deformations. 
The curves for different values of $\tau$ produce identical results when time is made dimensionless as $t/\tau$.  The predictions of VE model ($1-x_{VE}/{x_\infty}=e^{-t/\tau}$) are displayed as a dashed line. 
As it can be seen, the relaxation at early times
is independent of $\beta_0$ and controlled by $\tau$. 
This is followed by a slow relaxation dynamics, with the relaxation time appearing to increase with permeability, $\beta_\circ$. Note that $1-x/x_\infty \to 0$ as $t \to \infty$. 

Next, to study the effect of network compressibility we compute the relaxation dynamics for different values of $\nu$ and a constant $\beta_\circ=2$ in Fig.~\ref{fig1}(c). As it can be seen, the extent of deformation that follows a 
slow relaxation dynamics is reduced with increasing $\nu$. For $\nu=0.49$, which correspond to nearly incompressible network, the relaxation is almost entirely given by VE model. This observation suggest that the slow relaxation is induced by network compressibility. An inspection of Eq.~(\ref{eq4})
shows that, for a linear elastic network and a Newtonian fluid, it is a diffusion equation, with the diffusion coefficient given by $D_q=\tau^{-1} \alpha_\epsilon^{-2}$. The same equation and concept appears in Biot poroelasticity \citep{doi2009gel,cheng2016poroelasticity, detournay1993fundamentals}, where $D_q$ is sometimes referred to as generalized consolidation coefficient in soil mechanics literature.

Following this observation we rescale the time axis of Fig. \ref{fig1}(b) with $t/\tau_D$, where $\tau_D=a^2/D_q$ is the characteristic time-scale for the diffusion of network compressibility. Upon this rescaling, all the results collapse to a single line at longer time-scales (see the inset of Fig. \ref{fig1}(b)), which is in line with the idea that the slow relaxation is determined by the diffusion of network compressibility and the fluid pressure induced by that.
This rescaling did not result in the collapse of the data for different values of $\nu$ (see the inset of Fig. \ref{fig1}(c)), suggesting a more complex dependency of the dynamics on $\nu$.  
In figure \ref{fig1}(d), we study the effect of volume fraction of the network phase on probe dynamics in the range $0 < \phi \le 0.3$ for the choice of $\beta_\circ=2$ and $\nu=0.3$. The value of $\phi$ appears not to change the qualitative behavior and only has a minor quantitative effect on the overall dynamics. 
We observe the same weak dependencies for the displacement fields. As such, in the remainder of this paper we only present the results for $\phi\to 0$, which results in $\epsilon=0$, $\eta_\epsilon=\eta,\,\tau_\epsilon=\tau$ and  $\alpha_\epsilon=\alpha$.

Note that even though $\phi \to 0$, the fluid 
permeability can still be largely affected by the cytoskeletal network. To see this, consider 
the expression for permeability of a dilute fibrous
network in a Newtonian fluid by \cite{sangani1982slow}:
$(\beta_\circ a_f)^{2}=\frac{8\phi}{-\ln \phi -1.48}$,
where $a_f$ is the radius of the fiber, and $a_f \approx 12 \, \text{nm}$ for microtubules that have the largest radius among the cytoskeletal filaments. 
Rescaling this expression to the probe's radius --which is the scale of interest here-- gives 
$({\beta_\circ} a)^2= (\frac{a}{a_f})^2\frac{8\phi}{-\ln \phi -1.48}$. 
Assuming a spherical bead of radius $ a\sim 1 \, \mu\text{m}$, we see that even for very small volume fractions we have the likely  condition of $\phi (a/a_f)^2 \gg 1$, 
which makes $\beta_\circ a \gg 1$ resulting in substantial
reduction in fluid permeability by the network in the probe's length-scale. 

Based on these findings we provide a recipe for determining the constitutive parameters in an active MR setup such as the one used here, when the fluid is
Newtonian. If the fluid is also VE, the probe motion alone can only determine the ratio of $\tilde{\eta}{s}/\tilde{G}(s)$ and not the individual terms, and displacement fields will be needed to disentangle the timescales associated with each the fluid and the network. At very early times the network is hardly deformed and the probe velocity is determined 
by the fluid drag force: $U (t\to 0)={\mathrm{F}}_\circ/(6\pi \eta a)$. Measuring $U (t=0)$ can, thus, be used to compute $\eta$. At later intermediate times the dynamics is controlled by relaxation of shear modes  allowing us to determine $G(t)/\eta$ and, thus, $G(t)$. Having $\eta$ and $G(t)$, one can compute $\nu$ by using the steady-state displacement of the probe: 
${x}_{\infty}=\frac{{{\mathrm{F}}_{\circ}}}{6\pi G(t\to \infty) a} \frac{5-6\nu}{4(1-\nu)}$. Finally, $\beta_{\circ}$ can be determined by fitting Eq.~(\ref{eq23}) to the displacement vs time at longer timescales. 

\begin{figure}
    \centering
    \subfigure[]{\includegraphics[width=0.487\textwidth , height=5.4cm ]{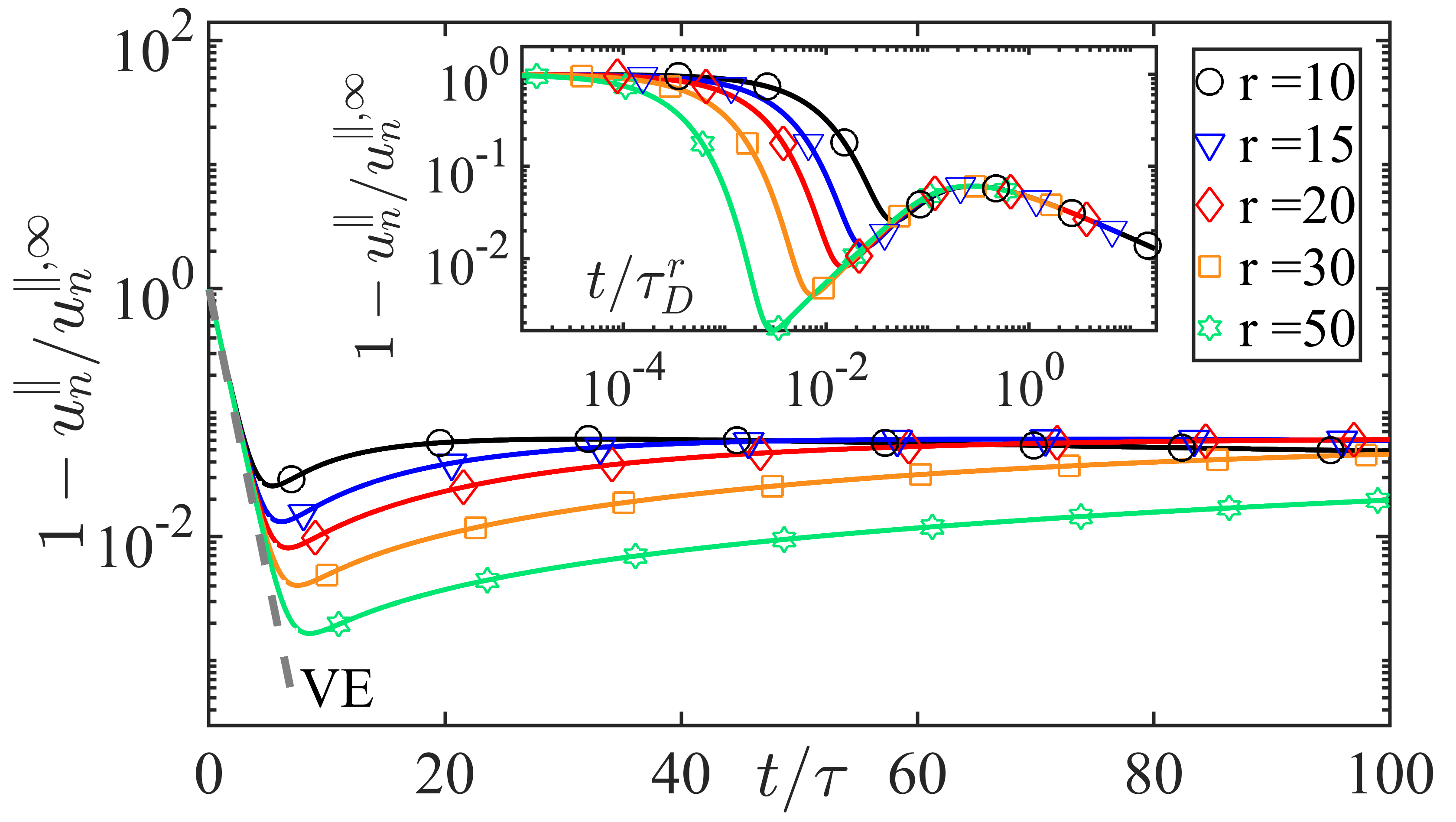}} \label{fig2a}
    \subfigure[]{\includegraphics[width=0.487\textwidth , height=5.4cm]{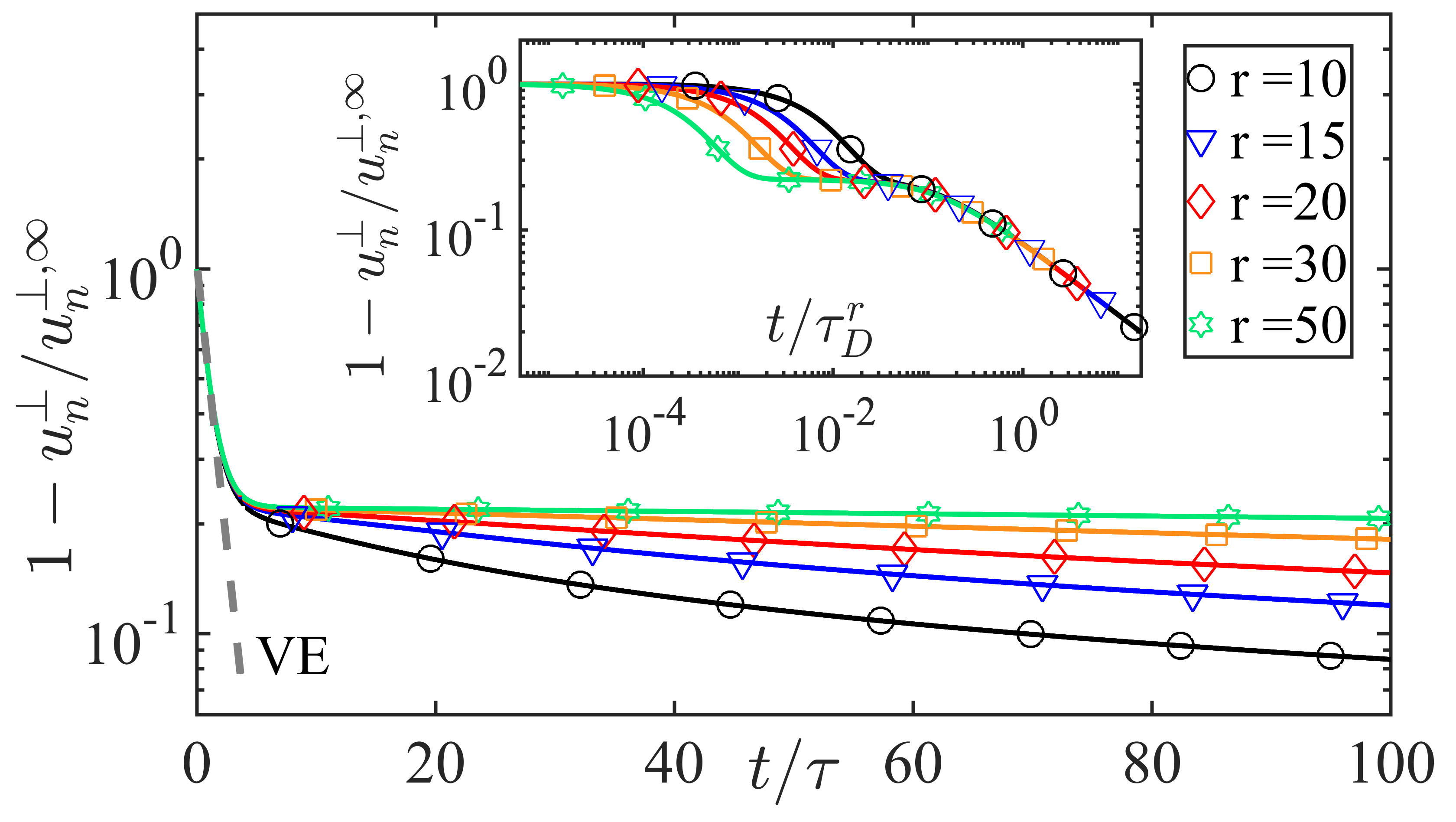}} \label{fig2b} 
    \caption{ The relaxation of network displacement fields in response to the moving probe in directions parallel (a) and perpendicular (b) to the applied force and 
        at different distances from the sphere, $1-u^{(\parallel,\perp)}_n/u^{(\parallel,\perp),\infty}_n$. The inset figures show the exact same results of the associated figures, when the time axis is scaled as
        $t/{\tau}_{D}^r$, where $\tau_D^r=r^2/D_q$ 
        is the time-scale for network compressibility field to diffuse distance $r$ away from the probe. Here we have taken ${\beta}_{\circ}=2$, $\phi \to 0$, and $\nu =0.3$.
    }
    \label{fig2}
\end{figure}

We now focus on the network displacement vs time at different, but all large, distances from the probe. Figure \ref{fig2}(a) shows displacement relaxation along the direction of the applied force, $1-u_{n}^{\parallel}(r)/u_{n}^{\parallel ,\infty}$, at different distances from the probe, for $\beta_\circ=2$, $\nu=0.3$ and $\phi\to 0$ (as for the rest of the results). Similar to the probe's dynamics, the relaxation at early times is
determined by the shear modes given by VE model (dashed line). For large $r/a$, the time-dependent $1-u_{n}^{\parallel}/u_{n}^{\parallel ,\infty}$ exhibit a slow relaxation dynamics with a local minimum (a minimum for $u_{n}^{\parallel}$) followed by what appears to be a local maximum, noting that $1-u_{n}^{\parallel}/u_{n}^{\parallel ,\infty} \to 0$ as $t\to \infty$. 
To test if the slow relaxation is again controlled to the diffusion of the stress associated with network compressibility, we scale the time axis with the diffusion time-scale for travelling the distance $r$: $\tau_D^{r}=r^2/D_q$. Again, we observe a nice collapse of the plots at longer times, which further confirms that slow relaxation dynamics in determined by the diffusion of network compressibility and its associated stresses. 

Figure \ref{fig2}(b) shows the displacement relaxation in the direction perpendicular to the applied force, $u_{n}^{\perp}$, vs time at different values of $r$. 
Unlike the displacements in parallel direction, we do not observe any local optimum in time. Aside form this difference the relaxation dynamics closely follows the behavior we observed in parallel direction. In Appendix \ref{appD}, we show the relaxation of fluid velocity vs time at different $r$. Unlike the network displacement dynamics,
the fluid velocity does not undergo a slow relaxation process and the relaxation dynamics is
dominated  by $\tau$. This difference can be explained by noting that the fluid, unlike the network, is incompressible; hence, its dynamics is determined by shear hydrodynamic modes and not the compression modes.  
A detailed study of the effect of $\beta$ and
$\nu$  on the relaxation dynamics of both network and fluid  displacement fields further confirms
our observations thus far, \textit{i.e.} for the network displacement fields the early/fast relaxation dynamics is controlled by $\tau$, and the slow relaxation is controlled by $\tau_D^r$, whereas
the relaxation of fluid velocity is determined only by $\tau$. 

\begin{figure}
    \centering
    \subfigure[]{\includegraphics[width=0.487\textwidth , height=5.4cm ]{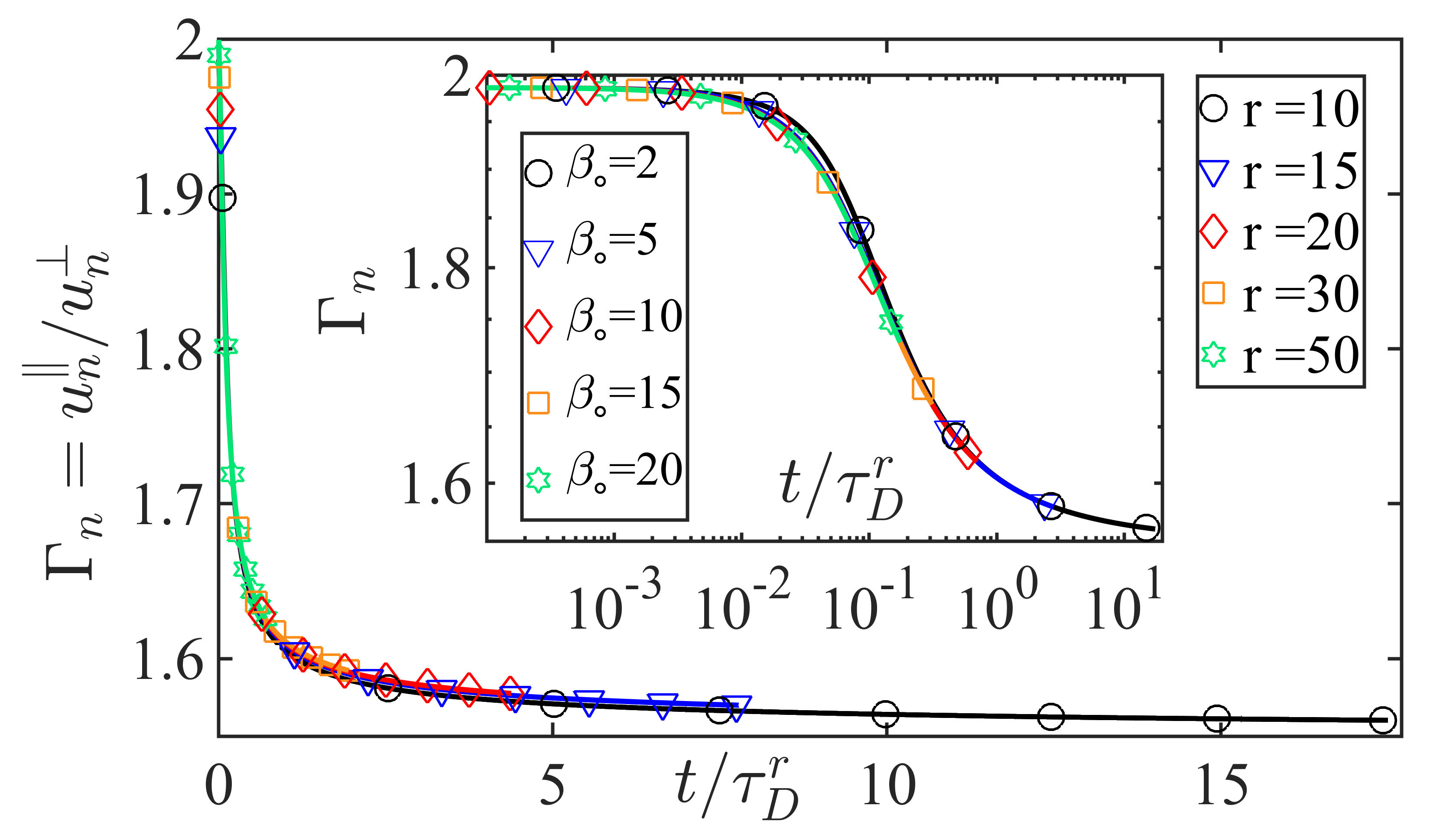}} \label{fig3a}
    \subfigure[]{\includegraphics[width=0.487\textwidth , height=5.4cm]{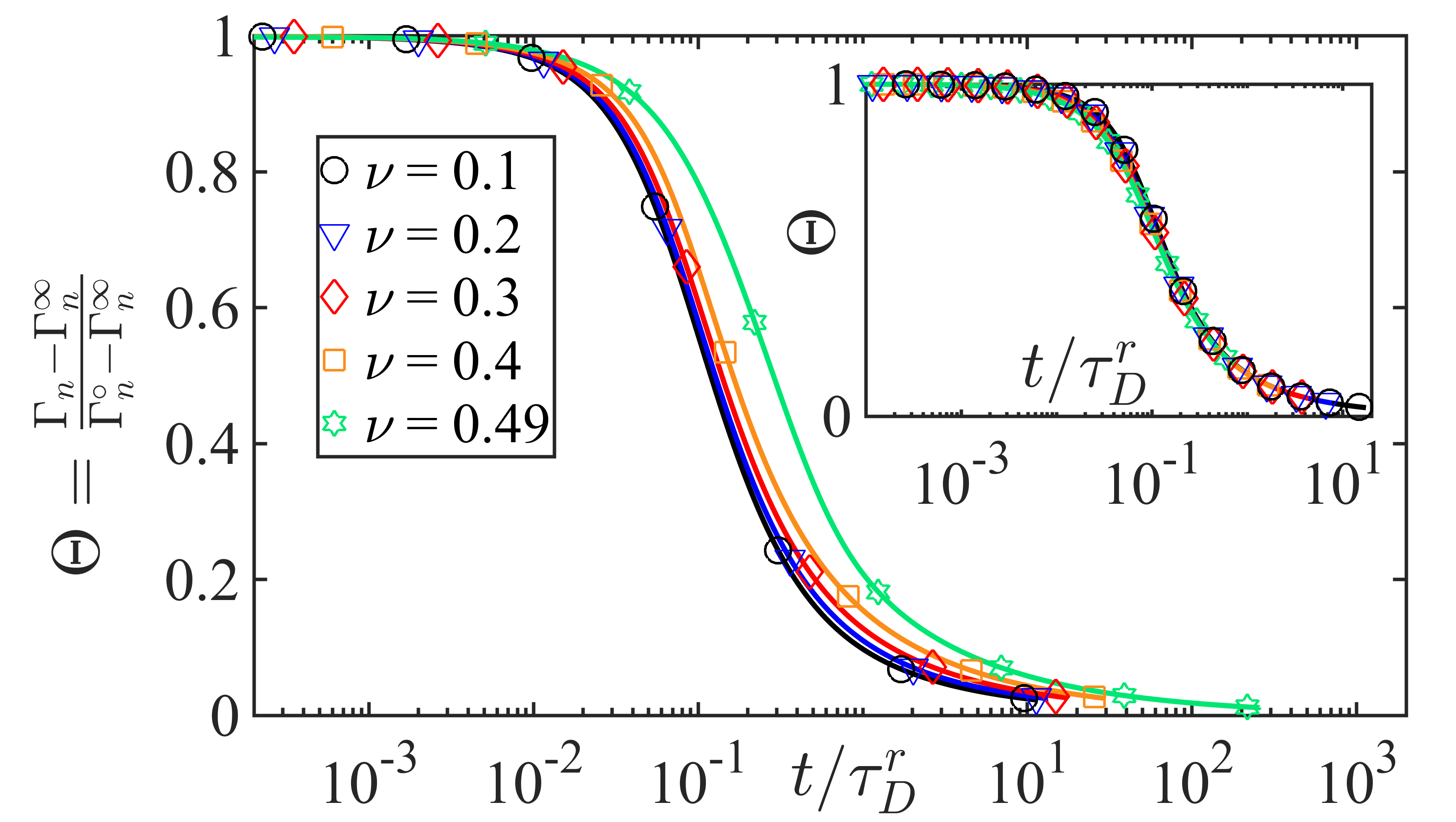}} \label{fig3b} 
    \caption{ (a) The ratio of parallel to perpendicular network displacements, $\Gamma_n=u^\parallel_n/u^\perp_n$, as a function of $t/\tau^D_r$. Main figure: different values of  $r$, taking $\beta_\circ=2$, and $\nu=0.3$. Inset: different $\beta_\circ$, taking $r=10$, $\nu=0.3$. 
    (b) Variations of $\Theta=\frac{\Gamma_n -\Gamma_{n}^{\infty}}{\Gamma_{n}^{\circ}-\Gamma_{n}^{\infty}}$ vs time for different Poisson rations taking $\beta_\circ=2$, and $r=10$. Inset: collection of all the data in figures 3(a-b), plotted as $\Theta$ vs $\frac{t}{\tau_D^r}$. All the data collapse into a single curve irrespective of the choice of $\nu$, $r$, $\beta_\circ$ and $\tau$.
    }
    \label{fig3}
\end{figure}

Another difference between VE and PE models appear in the ratio of parallel to perpendicular displacements. For VE materials, $u_\text{VE}^\parallel/u_\text{VE}^\perp=M_{12}^{\parallel}/M_{12}^{\perp}=2$, whereas for a linear elastic material the ratio is dependent on $\nu$ \citep{levine2000one}: $u_\text{E}^\parallel/u_\text{E}^\perp=\frac{4(1-\nu)}{3-4\nu}$, which changes from 
$2$ for $\nu=0.5$ to $\frac{8}{7}$ for $\nu=-1$.
For PE materials the ratio changes over time from $2$ at early times to $u_\text{E}^\parallel/u_\text{E}^\perp$ at long times. Figure \ref{fig3}(a) shows these
variations over time for $10\le r/a \le 50$ and $\beta_\circ=2$, while the inset shows the results $2\le \beta_\circ \le 20$ and $r/a=50$. In both plots $\nu=0.3$ and the 
time is made dimensionless by $\tau_D^r=r^2/D_q$.
As it can be seen, the results for different values of $r$ and $\beta_\circ$ all collapse to a single curve over the entire time domain. (See appendix \ref{appD} for plot of the ratio of parallel to perpendicular fluid displacement.) 
In figure \ref{fig3}(b) we study the effect of $\nu$ on the variations of $\Theta=\frac{\Gamma_n -\Gamma_{n}^\infty}{\Gamma_{n}^{\circ}-\Gamma_{n}^\infty}$ as a function of $t/\tau_D^r$ for different values of $\nu$, $\beta_\circ=2$  
and $r/a=10$, where $\Gamma_n (r,t)=u_n^{\parallel}/u_n^\perp$. 
The data collapse to a single curve.
The inset of figure \ref{fig3}(b) shows the cumulative $\Theta(t)$ vs 
$t/\tau_D^r$ data for different values of $r$, 
$\beta_\circ$ and $\nu$, collapsing into the same curve. All together, these results show that the relaxation dynamics of displacement (and mobility) ratios is entirely determined by the diffusion of network compressibility:  $\tau_D^r$. The experimental determination of this ratio in two-point MR provides another method to  compute $D_q$ and, thus, $\beta_\circ$. 

\section{Summary}
Increasing experimental evidence point to a PE model as the most accurate description 
of the mechanics of the cell cytoskeleton \citep{mogilner2018intracellular}. 
Biot's theory of poroelasticity has been used to 
describe these observations \citep{charras2005non, charras2008life}. 
Biot's theory neglects viscous stresses, which 
are crucial for determining the mechanical response of cytoskeletal assemblies ($\phi \ll 1$) at short and intermediate timescales. 
To overcome this limitation we include the viscous stresses, which modifies the fluid momentum equation from Darcy to Brinkman equation.
%
We present the first general solution
to these modified PE equations in spherical coordinates; the fluid phase is described as an incompressible linear viscoelastic fluid and the network phase is modeled as a compressible linear viscoelastic material. Similar to Stokes flow microhydrodynamics \citep{kim2013microhydrodynamics} and linear elasticity \citep{gurtin1973linear}, the linearity of the equations of motion allows developing a robust mathematical framework   
for describing the motion of inclusions moving in PE materials, and the interior displacements generated
in PE materials by the motion of outer boundaries.
The general solution provided here is one of several pieces of this framework.  

To demonstrate the utility of the general solution
we studied the dynamics of a rigid sphere moving in a PE material. Using the axisymmetric form of the general solution, we derived closed form solutions
for the sphere's response function (hydrodynamic resistance), and fluid and network displacement
fields. 
We showed that the displacement relaxation
of the rigid sphere and the network displacements
induced by it follow two distinct dynamics at short and long timescales. The dynamics at short timescales is
dominated by hydrodynamic shear modes
on the system, and, thus, can be captured using 
a VE model. The following slow relaxation dynamics 
could not be predicted using VE model; 
instead we showed that this dynamic is controlled by the diffusion timescale of network compressibility
and the response of the fluid pressure to these
local changes in network volume. 
These predictions are in agreement with the 
experimental observations, including the
slow distance-dependent relaxation that follows the initial fast relaxation observed by 
atomic force microscopy experiments of \cite{rosenbluth2008slow}. These observations are
in agreement with the results of \cite{charras2009animal} and \cite{moeendarbary2013cytoplasm}, and the 
experimental and computational observations of slow propagation of  hydrostatic pressure in 
blebs  \citep{charras2008life, charras2005non, strychalski2016intracellular}. 
Finally, we discussed how our results can
directly be used to analyze the results of single-particle and two-point MR and calculate fluid permeability 
in addition to the commonly measured 
viscoelastic properties of the system.

\appendix

\section{Solving ${\mathbf{v}}^{+}$ and ${\mathbf{v}}^{-}$ equations}\label{appA}
\subsection{General solution of  ${\mathbf{v}}^{+}$ equation }\label{appA1}
The equation for ${\mathbf{v}}^{+} $  is:
\begin{equation}
 {\nabla}^2 {\mathbf{v}}^{+}  -  \nabla (\nabla \cdot {\mathbf{v}}^{+}) = \nabla \Phi .
\end{equation} 
Using $\nabla \cdot  {\mathbf{v}}^{+} = -  (\varepsilon \tilde{\eta} +  \tilde{G} )  \, q$,  we can write the equation for ${\mathbf{v}}^{+}$ as:
\begin{equation}
{\nabla}^2 {\mathbf{v}}^{+} =  \nabla \Big( \Phi - (\varepsilon  \tilde{\eta} +  \tilde{G} ) \,q \Big) . \label{eq1appA}
\end{equation} 
We look for the solution of Eq.~\eqref{eq1appA} in the form $ {\mathbf{v}}^{+} =  {\mathbf{v}}_{\mathrm{h}}^{+} + \nabla \chi  $. Taking Laplacian of both side of this relation, and using Eq.~\eqref{eq1appA}, we see that $ {\mathbf{v}}_{\mathrm{h}}^{+}$ satisfies vector Laplace equation, ${\nabla}^2  {\mathbf{v}}_{\mathrm{h}}^{+} = \mathbf{0}$, and $\chi$ satisfies the following equation:
\be 
 {\nabla}^2 \chi = \Phi -  (\varepsilon  \tilde{\eta} +  \tilde{G} )  \,q  . \label{eq2appA}
\ee
The general solution of $ {\mathbf{v}}_{\mathrm{h}}^{+}$ is \cite{morse1953methods,arfken1999mathematical}:
 \be 
  {\mathbf{v}}_{\mathrm{h}}^{+} =   \sum_{\ell ,m}  \left[ {A}_{\ell ,m}^{\pm} {\mathbf{M}}_{\ell ,m}^{\pm} + {B}_{\ell ,m}^{\pm} {\mathbf{N}}_{\ell ,m}^{\pm} + {C}_{\ell ,m}^{\pm} {\mathbf{G}}_{\ell ,m}^{\pm}   \right] ,  \label{eq3appA}
 \ee
 where
\begin{subequations} \label{eq4appA}
\begin{align}
&  {\mathbf{M}}_{\ell ,m}^{\pm} = \begin{pmatrix}
r^{\ell} \\ r^{- \ell -1} \end{pmatrix} \sqrt{\ell (\ell +1)} {\mathbf{C}}_{\ell ,m}  ,  \label{eq4aappA} \\
& {\mathbf{N}}_{\ell ,m}^{\pm} = \begin{pmatrix}  
r^{\ell} \\ r^{- \ell -1} \end{pmatrix}    \begin{pmatrix}
(\ell +1) {\mathbf{P}}_{\ell +1,m} + \sqrt{(\ell +1)(\ell +2)} {\mathbf{B}}_{\ell +1,m} \\
- \ell {\mathbf{P}}_{\ell -1,m} + \sqrt{\ell (\ell -1)} {\mathbf{B}}_{\ell -1,m}
\end{pmatrix} ,  \label{eq4bappA} \\
 & {\mathbf{G}}_{\ell ,m}^{\pm} = \begin{pmatrix}
r^{\ell} \\ r^{- \ell -1} \end{pmatrix}  \begin{pmatrix}
- \ell {\mathbf{P}}_{\ell -1,m} + \sqrt{\ell (\ell -1)} {\mathbf{B}}_{\ell -1,m} \\
(\ell +1) {\mathbf{P}}_{\ell +1,m} + \sqrt{(\ell +1)(\ell +2)} {\mathbf{B}}_{\ell +1,m} 
\end{pmatrix} , \label{eq4cappA}  
\end{align}
\end{subequations}
and $ {\mathbf{P}}_{\ell ,m}$, $ {\mathbf{B}}_{\ell ,m}$, and $ {\mathbf{C}}_{\ell ,m}$ are defined in the main text. Using Laplace general solution for $\Phi$, and Helmholtz general solution for $q$, Eq.~\eqref{eq2appA} becomes:
\be 
{\nabla}^2 \chi = \sum_{\ell ,m}  \left[  D_{\ell ,m}^{\pm}   \begin{pmatrix}
r^{\ell} \\ r^{-\ell -1} \end{pmatrix}  -  (\varepsilon  \tilde{\eta} +  \tilde{G})  \,\, E_{\ell ,m}^{\pm}  \begin{pmatrix}
{\mathsf{i}}_{\ell } ({\alpha}_{\varepsilon} r) \\ {\mathsf{k}}_{\ell } ({\alpha}_{\varepsilon} r) 
\end{pmatrix} \right]  Y_{\ell m} (\theta ,\varphi) . \label{eq5appA}
\ee
The solution for $\chi$ is:
\be 
\chi = \sum_{\ell ,m} \left[ D_{\ell ,m}^{\pm}  \frac{r^2}{2} \begin{pmatrix}
\frac{1}{2\ell +3} r^{\ell} \\ \frac{1}{-2\ell +1} r^{-\ell -1} \end{pmatrix} - \frac{ \varepsilon  \tilde{\eta} +  \tilde{G} }{{\alpha}_{\varepsilon}^2}  E_{\ell ,m}^{\pm}  \begin{pmatrix}
{\mathsf{i}}_{\ell } ({\alpha}_{\varepsilon} r) \\  {\mathsf{k}}_{\ell } ({\alpha}_{\varepsilon} r) 
\end{pmatrix}       \right] Y_{\ell m}(\theta, \varphi) .    \label{eq6appA}
\ee
So, $\nabla \chi$ becomes:
\be
\begin{aligned}
& \displaystyle{ \nabla \chi =  \sum_{\ell ,m}  \frac{1}{2}   \begin{pmatrix}
r^{\ell +1} \\ r^{-\ell } \end{pmatrix} D_{\ell ,m}^{\pm} \begin{pmatrix} 
\frac{\ell +2}{2\ell +3} {\mathbf{P}}_{\ell ,m} + \frac{1}{2\ell +3} \sqrt{\ell (\ell +1)} {\mathbf{B}}_{\ell ,m} \\
\frac{\ell -1}{2\ell -1} {\mathbf{P}}_{\ell ,m} + \frac{1}{-2\ell +1} \sqrt{\ell (\ell +1)} {\mathbf{B}}_{\ell ,m} 
\end{pmatrix}   } &&    \\ 
& \displaystyle{ \qquad\qquad\qquad  - \sum_{\ell ,m} \frac{ \varepsilon  \tilde{\eta} +  \tilde{G} }{{\alpha}_{\varepsilon}^2} E_{\ell ,m}^{\pm} \left[ \frac{\mathrm{d}}{\mathrm{d} r}  \begin{pmatrix}
{\mathsf{i}}_{\ell } ({\alpha}_{\varepsilon} r) \\  {\mathsf{k}}_{\ell } ({\alpha}_{\varepsilon} r) 
\end{pmatrix}  {\mathbf{P}}_{\ell ,m}+ \frac{1}{r} \begin{pmatrix}
{\mathsf{i}}_{\ell } ({\alpha}_{\varepsilon} r) \\  {\mathsf{k}}_{\ell } ({\alpha}_{\varepsilon} r) 
\end{pmatrix} \sqrt{\ell(\ell +1)} {\mathbf{B}}_{\ell ,m}  \right] .  }  &&    \label{eq7appA}
\end{aligned} 
\ee
\\
Therefore, the solution for ${\mathbf{v}}^{+}$ becomes:
\be
\begin{aligned}
& {\mathbf{v}}^{+} = \sum_{\ell ,m}     \begin{pmatrix}
r^{\ell} \\ r^{-\ell -1}
\end{pmatrix} \Bigg\{  A_{\ell ,m}^{\pm} \sqrt{\ell (\ell +1)} {\mathbf{C}}_{\ell ,m} 
+  B_{\ell ,m}^{\pm}      \begin{pmatrix}
(\ell +1) {\mathbf{P}}_{\ell +1,m} + \sqrt{(\ell +1)(\ell +2)} {\mathbf{B}}_{\ell +1,m} \\
- \ell {\mathbf{P}}_{\ell -1,m} + \sqrt{\ell (\ell -1)} {\mathbf{B}}_{\ell -1,m}
\end{pmatrix}  &&  \\
& \qquad +   C_{\ell ,m}^{\pm}   \begin{pmatrix}
- \ell {\mathbf{P}}_{\ell -1,m} + \sqrt{\ell (\ell -1)} {\mathbf{B}}_{\ell -1,m} \\
(\ell +1) {\mathbf{P}}_{\ell +1,m} + \sqrt{(\ell +1)(\ell +2)} {\mathbf{B}}_{\ell +1,m} 
\end{pmatrix}    \Bigg\}  +  \frac{1}{2}   \begin{pmatrix}
r^{\ell +1} \\ r^{-\ell } \end{pmatrix} D_{\ell ,m}^{\pm} \begin{pmatrix} 
\frac{\ell +2}{2\ell +3} {\mathbf{P}}_{\ell ,m} + \frac{1}{2\ell +3} \sqrt{\ell (\ell +1)} {\mathbf{B}}_{\ell ,m} \\
\frac{\ell -1}{2\ell -1} {\mathbf{P}}_{\ell ,m} + \frac{1}{-2\ell +1} \sqrt{\ell (\ell +1)} {\mathbf{B}}_{\ell ,m} 
\end{pmatrix}    &&  \\
 & \qquad\qquad\qquad\qquad  \displaystyle{ \qquad  - \frac{ \varepsilon  \tilde{\eta} +  \tilde{G} }{{\alpha}_{\varepsilon}^2} \,  E_{\ell ,m}^{\pm} \left[   \frac{\mathrm{d}}{\mathrm{d} r}  \begin{pmatrix}
{\mathsf{i}}_{\ell } ({\alpha}_{\varepsilon} r) \\  {\mathsf{k}}_{\ell } ({\alpha}_{\varepsilon} r) 
\end{pmatrix} {\mathbf{P}}_{\ell ,m}+ \frac{1}{r} \begin{pmatrix}
{\mathsf{i}}_{\ell } ({\alpha}_{\varepsilon} r) \\  {\mathsf{k}}_{\ell } ({\alpha}_{\varepsilon} r) 
\end{pmatrix}  \sqrt{\ell(\ell +1)}  {\mathbf{B}}_{\ell ,m}  \right] .  }                     &&    \label{eq8appA}
\end{aligned} 
\ee
\\
Setting $\nabla \cdot  {\mathbf{v}}^{+} = -  (\varepsilon  \tilde{\eta} +  \tilde{G} )  \, q$, gives:
\be 
{D}_{\ell ,m}^{+} = (\ell +1) (2\ell +3) {C}_{\ell +1 ,m}^{+} , \qquad\qquad\qquad\qquad\qquad
 {D}_{\ell ,m}^{-} = \ell (2\ell -1) {C}_{\ell -1 ,m}^{-} ,    \label{eq9appA}
\ee
and we get Eq.~\eqref{eq9} in the main text.

\subsection{General solution of  ${\mathbf{v}}^{-}$ equation}\label{appA2}
 The equation for ${\mathbf{v}}^{-}$ is:
\be 
{\nabla}^2  {\mathbf{v}}^{-} + \frac{{\gamma}_{\varepsilon}}{1- \varepsilon} \nabla (\nabla \cdot  {\mathbf{v}}^{-} ) - {\beta}^2 {\mathbf{v}}^{-} = \frac{1}{{ \tilde{\eta}}_{\varepsilon}} \nabla \Phi.  \label{eq10appA}
\ee
We look for the solution of Eq.\eqref{eq10appA} in the form $ {\mathbf{v}}^{-} =   {\mathbf{v}}_{\mathrm{h}}^{-} +  {\mathbf{v}}_{\mathrm{p}}^{-}   $. For the homogeneous part we assume $ {\mathbf{v}}_{\mathrm{h}}^{-} =  
 {\mathbf{v}}_{\mathrm{T}}^{-} +  {\mathbf{v}}_{\mathrm{L}}^{-} $, where $\nabla \cdot  {\mathbf{v}}_{\mathrm{T}}^{-} =0$, and $\nabla \times  {\mathbf{v}}_{\mathrm{L}}^{-} = \mathbf{0} $.
 Inserting this ansatz to  ${\mathbf{v}}^{-}$ equation, we get:
\be 
{\nabla}^2  {\mathbf{v}}_{\mathrm{T}}^{-} + {\nabla}^2  {\mathbf{v}}_{\mathrm{L}}^{-} + \frac{{\gamma}_{\varepsilon}}{1-\varepsilon} \nabla \Big( \nabla \cdot {\mathbf{v}}_{\mathrm{L}}^{-} \Big) - {\beta}^2 {\mathbf{v}}_{\mathrm{T}}^{-}  - {\beta}^2 {\mathbf{v}}_{\mathrm{L}}^{-} = \mathbf{0} , \label{eq11appA}
\ee
which could be fulfilled by requiring that:
\begin{subequations} \label{eq12appA}
\begin{align}
 & {\nabla}^2  {\mathbf{v}}_{\mathrm{T}}^{-}   - {\beta}^2 {\mathbf{v}}_{\mathrm{T}}^{-} = \mathbf{0} ,                     \label{eq12aappA} \\
 &  {\nabla}^2  {\mathbf{v}}_{\mathrm{L}}^{-} + \frac{{\gamma}_{\varepsilon}}{1-\varepsilon} \nabla \Big( \nabla \cdot {\mathbf{v}}_{\mathrm{L}}^{-} \Big) - {\beta}^2 {\mathbf{v}}_{\mathrm{L}}^{-} =         \mathbf{0}                               . \label{eq12bappA}
\end{align}
\end{subequations}
Since, $\nabla \Big( \nabla \cdot {\mathbf{v}}_{\mathrm{L}}^{-} \Big) =  {\nabla}^2  {\mathbf{v}}_{\mathrm{L}}^{-} $, we can write:
\begin{subequations} \label{eq13appA}
\begin{align}
 & {\nabla}^2  {\mathbf{v}}_{\mathrm{T}}^{-}   = {\beta}^2 {\mathbf{v}}_{\mathrm{T}}^{-} ,    \label{eq13aappA}\\
 & {\nabla}^2  {\mathbf{v}}_{\mathrm{L}}^{-} = {\alpha_{\varepsilon}}^2 {\mathbf{v}}_{\mathrm{L}}^{-}  ,  \label{eq3bappA}
\end{align}
\end{subequations}
where $ {\alpha_{\varepsilon}}^2 =   \frac{{\beta}^2 (1- \varepsilon)}{1- \varepsilon + {\gamma}_{\varepsilon}}$.
The general solution for ${\mathbf{v}}_{\mathrm{T}}^{-}$, and ${\mathbf{v}}_{\mathrm{L}}^{-}$ are \cite{ben1968eigenvector}:
\begin{subequations} \label{eq14appA}
\begin{align}
& {\mathbf{v}}_{\mathrm{T}}^{-} = \sum_{\ell ,m} {A'}_{\ell ,m}^{\pm} {\mathbf{M '} }_{\ell ,m}^{\pm} + {B '}_{\ell ,m}^{\pm} {\mathbf{N '} }_{\ell ,m}^{\pm} ,  \label{eq14aappA}  \\
 & {\mathbf{v}}_{\mathrm{L}}^{-} = \sum_{\ell ,m} {C '}_{\ell ,m}^{\pm} {\mathbf{L} }_{\ell ,m}^{\pm} , \label{eq14bappA}
\end{align}
\end{subequations}
where $ {\mathbf{M '} }_{\ell ,m}^{\pm} = \nabla \times \left( \mathbf{r} \, {\psi}_{\mathrm{T}, \ell ,m}^{\pm} \right) $, ${\mathbf{N '} }_{\ell ,m}^{\pm} = \frac{1}{\beta}  \nabla \times \nabla \times \left( \mathbf{r} \, {\psi}_{\mathrm{T}, \ell ,m}^{\pm} \right)$, and ${\mathbf{L} }_{\ell ,m}^{\pm} = \frac{1}{{\alpha}_{\varepsilon}} \nabla {\psi}_{\mathrm{L}, \ell ,m}^{\pm} $.
Here, $ {\psi}_{\mathrm{T}, \ell ,m}^{\pm} $, and $ {\psi}_{\mathrm{L}, \ell ,m}^{\pm} $ satisfy scalar Helmholtz equation ${\nabla}^2 {\psi}_{\mathrm{T}, \ell ,m}^{\pm} - {\beta}^2 {\psi}_{\mathrm{T}, \ell ,m}^{\pm} =0 $, and ${\nabla}^2 {\psi}_{\mathrm{L}, \ell ,m}^{\pm} - {{\alpha}_{\varepsilon}}^2 {\psi}_{\mathrm{L}, \ell ,m}^{\pm} =0$, respectively. So:
\be 
{\mathbf{v}}_{\mathrm{h}}^{-} =  \sum_{\ell ,m} {A'}_{\ell ,m}^{\pm} {\mathbf{M '} }_{\ell ,m}^{\pm} + {B '}_{\ell ,m}^{\pm} {\mathbf{N '} }_{\ell ,m}^{\pm} + {C '}_{\ell ,m}^{\pm} {\mathbf{L} }_{\ell ,m}^{\pm} , \label{eq15appA}
\ee
where
\begin{subequations} \label{eq16appA}
\begin{align}
& {\mathbf{M '} }_{\ell ,m}^{\pm}  =   \begin{pmatrix}
{\mathsf{i}}_{\ell } (\beta r) \\  {\mathsf{k}}_{\ell } (\beta r) 
\end{pmatrix} \sqrt{\ell (\ell +1)} {\mathbf{C}}_{\ell ,m}     ,     \label{eq16aappA} \\
& {\mathbf{N '} }_{\ell ,m}^{\pm} = \ell (\ell +1) {\mathbf{P}}_{\ell ,m} \frac{1}{\beta r}  \begin{pmatrix}
{\mathsf{i}}_{\ell } (\beta r) \\  {\mathsf{k}}_{\ell } (\beta r) 
\end{pmatrix} +   \sqrt{\ell (\ell +1)}  {\mathbf{B}}_{\ell ,m}  \left\lbrace \frac{\mathrm{d}}{\mathrm{d} (\beta r)}   \begin{pmatrix}
{\mathsf{i}}_{\ell } (\beta r) \\  {\mathsf{k}}_{\ell } (\beta r) 
\end{pmatrix}  + \frac{1}{\beta r}  \begin{pmatrix}
{\mathsf{i}}_{\ell } (\beta r) \\  {\mathsf{k}}_{\ell } (\beta r) 
\end{pmatrix}  \right\rbrace   ,   \label{eq16bappA} \\
& {\mathbf{L} }_{\ell ,m}^{\pm} =    \frac{\mathrm{d}}{\mathrm{d} ({\alpha}_{\varepsilon} r)}   \begin{pmatrix}
{\mathsf{i}}_{\ell } ({\alpha}_{\varepsilon} r) \\  {\mathsf{k}}_{\ell } ( {\alpha}_{\varepsilon} r) 
\end{pmatrix}  {\mathbf{P}}_{\ell ,m} +  \sqrt{\ell (\ell +1)}  {\mathbf{B}}_{\ell ,m}     \frac{1}{ {\alpha}_{\varepsilon} r}  \begin{pmatrix}
{\mathsf{i}}_{\ell } ( {\alpha}_{\varepsilon} r) \\  {\mathsf{k}}_{\ell } ( {\alpha}_{\varepsilon} r) 
\end{pmatrix}   .      \label{eq16cappA}
\end{align}
\end{subequations}
\\
Next, we seek for the particular solution of the form $ {\mathbf{v}}_{\mathrm{p}}^{-} =  \nabla \chi $. Inserting this ansatz to Eq.\eqref{eq10appA}, we get:
\be
 {\nabla}^2 \chi - {\beta}^2 \chi =  \frac{1}{{ \tilde{\eta}}_{\varepsilon}}   \Phi - \gamma_{\varepsilon} \, q . \label{eq17appA}
\ee
Assuming $\chi = {\chi}_1 + {\chi}_2$, where ${\chi}_1$ and ${\chi}_2$ satisfy equations $ {\nabla}^2 {\chi}_1 - {\beta}^2 {\chi}_1 = - \gamma_{\varepsilon} \, q$, and $ {\nabla}^2 {\chi}_2 - {\beta}^2 {\chi}_2 =   \frac{1}{{ \tilde{\eta}}_{\varepsilon}}  \Phi $, we get solutions ${\chi}_1 = - \frac{\gamma_{\varepsilon}}{ {\alpha}_{\varepsilon}^2 - {\beta}^2} \, q  $, and $ {\chi}_2 = - \frac{1}{{\beta}^2}  \frac{1}{{ \tilde{\eta}}_{\varepsilon}}  \Phi $. 
Using general solution of Laplace for $\Phi$, and modified Helmholtz equation for $q$, we get:
\be 
\chi = {\chi}_1 + {\chi}_2 = \sum_{\ell ,m}  \left[ - \frac{1}{{\beta}^2}  \frac{1}{{ \tilde{\eta}}_{\varepsilon}}  D_{\ell ,m}^{\pm} \begin{pmatrix} 
r^{\ell} \\ r^{-\ell -1} \end{pmatrix}  - \frac{\gamma_{\varepsilon}}{ {\alpha}_{\varepsilon}^2 - {\beta}^2} E_{\ell ,m}^{\pm} \begin{pmatrix}
{\mathsf{i}}_{\ell } ({\alpha}_{\varepsilon} r) \\ {\mathsf{k}}_{\ell } ({\alpha}_{\varepsilon} r) 
\end{pmatrix} \right] Y_{\ell m} (\theta ,\varphi).  \label{eq19appA}
\ee
So, the particular solution is:
\be
\begin{aligned} 
& {\mathbf{v}}_{\mathrm{p}}^{-} = \nabla \chi = \sum_{\ell ,m} \Bigg\{     - \frac{1}{{\beta}^2}  \frac{1}{{ \tilde{\eta}}_{\varepsilon}}  {D}_{\ell ,m}^{\pm} \begin{pmatrix}
r^{\ell -1} \\ r^{-\ell -2}
\end{pmatrix}  \begin{pmatrix}
\ell {\mathbf{P}}_{\ell ,m} + \sqrt{\ell (\ell +1)} {\mathbf{B}}_{\ell ,m} \\
-(\ell +1) {\mathbf{P}}_{\ell ,m} + \sqrt{\ell (\ell +1)} {\mathbf{B}}_{\ell ,m}
\end{pmatrix}  \qquad\qquad\qquad\qquad \qquad &&  \\
&  \qquad\qquad \qquad   - \frac{\gamma_{\varepsilon}}{ {\alpha}_{\varepsilon}^2 - {\beta}^2}  {E}_{\ell ,m}^{\pm}  \left[ \frac{\mathrm{d}}{\mathrm{d} r}  \begin{pmatrix}
{\mathsf{i}}_{\ell } ( {\alpha}_{\varepsilon} r) \\  {\mathsf{k}}_{\ell } ( {\alpha}_{\varepsilon} r) 
\end{pmatrix} {\mathbf{P}}_{\ell ,m}    + \frac{1}{r}    \begin{pmatrix}
{\mathsf{i}}_{\ell } ( {\alpha}_{\varepsilon} r) \\  {\mathsf{k}}_{\ell } ( {\alpha}_{\varepsilon} r) 
\end{pmatrix} \sqrt{\ell (\ell +1)}  {\mathbf{B}}_{\ell ,m}         \right]  \Bigg\}  .  && \label{eq20appA}
\end{aligned}
\ee
Therefore, the general solution for ${\mathbf{v}}^{-}$ becomes:
\be 
\begin{aligned}
&  {\mathbf{v}}^{-} =    \sum_{\ell ,m} \Bigg\{      {A'}_{\ell ,m}^{\pm}    \begin{pmatrix}
{\mathsf{i}}_{\ell } (\beta r) \\  {\mathsf{k}}_{\ell } (\beta r) 
\end{pmatrix} \sqrt{\ell (\ell +1)} {\mathbf{C}}_{\ell ,m}  +  {B '}_{\ell ,m}^{\pm}  \Bigg[  \ell (\ell +1) {\mathbf{P}}_{\ell ,m} \frac{1}{\beta r}  \begin{pmatrix}
{\mathsf{i}}_{\ell } (\beta r) \\  {\mathsf{k}}_{\ell } (\beta r) 
\end{pmatrix} +  +  \sqrt{\ell (\ell +1)}  {\mathbf{B}}_{\ell ,m}  \Bigg(    \frac{\mathrm{d}}{\mathrm{d} (\beta r)}   \begin{pmatrix}
{\mathsf{i}}_{\ell } (\beta r) \\  {\mathsf{k}}_{\ell } (\beta r) 
\end{pmatrix}    \qquad\qquad    && \\
& \qquad\qquad\qquad  + \frac{1}{\beta r}  \begin{pmatrix}
{\mathsf{i}}_{\ell } (\beta r) \\  {\mathsf{k}}_{\ell } (\beta r) 
\end{pmatrix}  \Bigg)   \Bigg]   +  {C '}_{\ell ,m}^{\pm}   \Bigg[   \frac{\mathrm{d}}{\mathrm{d} ({\alpha}_{\varepsilon} r)}   \begin{pmatrix}
{\mathsf{i}}_{\ell } ({\alpha}_{\varepsilon} r) \\  {\mathsf{k}}_{\ell } ( {\alpha}_{\varepsilon} r) 
\end{pmatrix}  {\mathbf{P}}_{\ell ,m} +  \sqrt{\ell (\ell +1)}  {\mathbf{B}}_{\ell ,m}     \frac{1}{ {\alpha}_{\varepsilon} r}  \begin{pmatrix}
{\mathsf{i}}_{\ell } ( {\alpha}_{\varepsilon} r) \\  {\mathsf{k}}_{\ell } ( {\alpha}_{\varepsilon} r) 
\end{pmatrix}           \Bigg]       && \\
& \qquad  \qquad\qquad    - \frac{1}{{\beta}^2}  \frac{1}{{ \tilde{\eta}}_{\varepsilon}}  {D}_{\ell ,m}^{\pm} \begin{pmatrix}
r^{\ell -1} \\ r^{-\ell -2}
\end{pmatrix}  \begin{pmatrix}
\ell {\mathbf{P}}_{\ell ,m} + \sqrt{\ell (\ell +1)} {\mathbf{B}}_{\ell ,m} \\
-(\ell +1) {\mathbf{P}}_{\ell ,m} + \sqrt{\ell (\ell +1)} {\mathbf{B}}_{\ell ,m}
\end{pmatrix}   - \frac{\gamma_{\varepsilon}}{ {\alpha}_{\varepsilon}^2 - {\beta}^2}  {E}_{\ell ,m}^{\pm}  \Bigg[ \frac{\mathrm{d}}{\mathrm{d} r}  \begin{pmatrix}
{\mathsf{i}}_{\ell } ( {\alpha}_{\varepsilon} r) \\  {\mathsf{k}}_{\ell } ( {\alpha}_{\varepsilon} r) 
\end{pmatrix} {\mathbf{P}}_{\ell ,m}         && \\
 & \qquad\qquad  \qquad   + \frac{1}{r}    \begin{pmatrix}
{\mathsf{i}}_{\ell } ( {\alpha}_{\varepsilon} r) \\  {\mathsf{k}}_{\ell } ( {\alpha}_{\varepsilon} r) 
\end{pmatrix} \sqrt{\ell (\ell +1)}  {\mathbf{B}}_{\ell ,m}         \Bigg]                 \Bigg\} . && \label{eq21appA}
\end{aligned}
\ee
Using the fact that $\displaystyle{  \frac{{\gamma}_{\varepsilon}}{{\beta}^2 -  {\alpha}_{\varepsilon}^2} = \frac{1-\varepsilon}{ {\alpha}_{\varepsilon}^2}    } $, and $\nabla \cdot  {\mathbf{v}}^{-}  = (1- \varepsilon )  q $, we get $ {C '}_{\ell ,m}^{\pm} =0$, and recover Eq.~\eqref{eq10} in the main text.

\subsection{Some relations between vector spherical functions}\label{appA3}
\begin{subnumcases}{}
 \nabla \times \Big(  f(r) {\mathbf{P}}_{\ell ,m}  \Big) = \frac{f(r)}{r} \sqrt{\ell (\ell +1)}  {\mathbf{C}}_{\ell ,m} , \qquad\qquad\qquad\qquad \nabla \cdot \Big(  f(r) {\mathbf{P}}_{\ell ,m}  \Big) = \Big( \frac{2}{r} f(r) +  \frac{\mathrm{d}f}{\mathrm{d}r}   \Big) Y_{\ell ,m} (\theta ,\varphi) , \nn  \\
  \nabla \times \Big(  f(r) {\mathbf{B}}_{\ell ,m}  \Big) = - \Big( \frac{f(r)}{r} + \frac{\mathrm{d}f}{\mathrm{d}r}  \Big)   {\mathbf{C}}_{\ell ,m} , \qquad\qquad\qquad\qquad  \nabla \cdot \Big(  f(r) {\mathbf{B}}_{\ell ,m}  \Big) = - \sqrt{\ell (\ell +1)} \frac{f(r)}{r} Y_{\ell ,m} (\theta ,\varphi) , \nn  \\
  \nabla \times \Big(  f(r) {\mathbf{C}}_{\ell ,m}  \Big) = \frac{f(r)}{r} \sqrt{\ell (\ell +1)}  {\mathbf{P}}_{\ell ,m} +  \Big( \frac{f(r)}{r} + \frac{\mathrm{d}f}{\mathrm{d}r}  \Big) {\mathbf{B}}_{\ell ,m} , \qquad\qquad \nabla \cdot \Big(  f(r) {\mathbf{C}}_{\ell ,m}  \Big) = 0 .    \nn
\end{subnumcases}

\section{Fluid and network Stress fields  }\label{appB}
Fluid stress components are:
\be 
\begin{aligned} 
& {\tilde{\sigma}}_{rr}^{\mathrm{f}} = -\tilde{p} (1-\phi)+ 2  \tilde{\eta} \frac{\partial  \tilde{v}_{r,\mathrm{f}}}{\partial r}  &&  \\
& \qquad = \frac{3a \,  \tilde{\eta} }{2 {\beta}^2 (1-\varepsilon) r^4} \frac{ \tilde{U}(s)}{{\Delta}_{\varepsilon}}   \Bigg[ 24 \varepsilon (1-\varepsilon)  (\frac{1+\tau}{\tau}) (1+a {\alpha}_{\varepsilon}) + 4 a^2 {\alpha}_{\varepsilon}^2 (1-\varepsilon) (2\varepsilon \tau -1 + 3\varepsilon)    -12 \frac{{\alpha}_{\varepsilon}^2}{{\beta}^2} {(1-\varepsilon)}^2 (1+a\beta)     &&  \\
& \qquad\qquad + (\varepsilon -1) \frac{{ \tilde{\eta}}_{\varepsilon}}{ \tilde{\eta}} \Big[ 2 a^2 \tau (\varepsilon -1) + \Big( (3-2\varepsilon)\tau +1  \Big) r^2  \Big]  \Big[ 2 {\beta}^2 (1+ a {\alpha}_{\varepsilon}) + {\alpha}_{\varepsilon}^2 (1+ a\beta + a^2 {\beta}^2)  \Big]        \\
& \qquad\qquad   +  4 \frac{{\alpha}_{\varepsilon}^2}{{\beta}^2} {(\varepsilon -1)}^2 (3+ 3\beta r + {\beta}^2 r^2) e^{-\beta (r-a)} + 24 \varepsilon (\varepsilon -1) (1+\tau) (1+ {\alpha}_{\varepsilon} r) e^{- {\alpha}_{\varepsilon} (r-a)}   +  2 {\alpha}_{\varepsilon}^2 r^2 \Big( \varepsilon \tau (-5+6\varepsilon)     &&  \\ 
 & \qquad\qquad + \frac{\tilde{\lambda}}{\tilde{G}}  + \frac{1}{\tau} ( \frac{\tilde{\lambda}}{\tilde{G}}+2) + (2-5\varepsilon + 6 {\varepsilon}^2)  \Big) e^{- {\alpha}_{\varepsilon} (r-a)}  + 2   {\alpha}_{\varepsilon}^3 r^3 (1+ \frac{1}{\tau}) \Big( \varepsilon \tau (-1 + 2\varepsilon)  + \frac{\tilde{\lambda}}{\tilde{G}}+2  \Big) e^{- {\alpha}_{\varepsilon} (r-a)}       \Bigg] \cos\theta ,     && \label{eq24appC}
\end{aligned}
\ee
\be 
\begin{aligned} 
 {\tilde{\sigma}}_{r\theta}^{\mathrm{f}} &= \tilde{\eta} \Big( \frac{1}{r} \frac{\partial  \tilde{v}_{r,\mathrm{f}}}{\partial \theta} + \frac{\partial  \tilde{v}_{\theta , \mathrm{f}}}{\partial r} - \frac{ \tilde{v}_{\theta , \mathrm{f}}}{r}   \Big)  &&  \\
& \qquad =  \frac{3a \,  \tilde{\eta} }{2 {\beta}^2 r^4} \frac{ \tilde{U}(s)}{{\Delta}_{\varepsilon}}   \Bigg[  12 \varepsilon (1+\tau) (1+ a {\alpha}_{\varepsilon}) -2 a^2 {\beta}^2 (\varepsilon -1) {\tau}_{\varepsilon}  (1+ a {\alpha}_{\varepsilon})  &&  \\
& \qquad\qquad  - \frac{{\alpha}_{\varepsilon}^2}{{\beta}^2} \Big[  6(1-\varepsilon) (1+ a\beta) - 2 a^2 {\beta}^2 (-1 +\varepsilon (3+2\tau)) + a^2 {\beta}^2 {\tau}_{\varepsilon} (\varepsilon -1) (1+ a\beta + a^2 {\beta}^2)      \Big]     &&  \\
& \qquad\qquad   + \frac{{\alpha}_{\varepsilon}^2}{{\beta}^2} (1-\varepsilon)(6+ 6\beta r + 3 {\beta}^2 r^2 + {\beta}^3 r^3)  e^{-\beta (r-a)}  -4\varepsilon (1+\tau) (3+ 3{\alpha}_{\varepsilon} r + {\alpha}_{\varepsilon}^2 r^2) e^{- {\alpha}_{\varepsilon} (r-a)}   \Bigg] \sin\theta . && \label{eq25appC}
\end{aligned}
\ee
\\
Network stress components are:
\be 
\begin{aligned} 
& {\tilde{\sigma}}_{rr}^{\mathrm{n}} =  ( \tilde{\lambda} + 2 \tilde{G})  \frac{\partial  \tilde{v}_{n,r} }{\partial r} +  2  \tilde{\lambda}  \frac{1}{r} \Big(  \tilde{v}_{n,r} +  \tilde{v}_{n, \theta} \cot\theta   \Big)     - \phi \tilde{p} &&  \\
& \qquad =   \frac{3a \,  \tilde{G} }{2 {\beta}^2 (1-\varepsilon) r^4} \frac{ \tilde{U}(s)}{{\Delta}_{\varepsilon}}   \Bigg[  24 (1-\varepsilon) (1+\tau) (1+ {\alpha}_{\varepsilon} r) + (\varepsilon -1) {\tau}_{\varepsilon}          \Big[ 2 a^2 (\varepsilon -1) - \Big(-2 + (3+\tau)\varepsilon  \Big) r^2    \Big]     \Big[  2 {\beta}^2 (1+ a {\alpha}_{\varepsilon})  &&  \\
& \qquad \qquad   + {\alpha}_{\varepsilon}^2 (1+ a\beta + a^2 {\beta}^2)  \Big]  + 4 \frac{{\alpha}_{\varepsilon}^2}{{\beta}^2} (\varepsilon -1) \Big[ \tau \Big( 3 (\varepsilon -1)(1+ a \beta) + (\varepsilon -3) a^2 {\beta}^2  \Big)  -2 a^2 {\beta}^2  \Big]         &&  \\
& \qquad \qquad    + 24 (\varepsilon -1) (1+\tau) (1+ {\alpha}_{\varepsilon} r) e^{- {\alpha}_{\varepsilon} (r-a)}       - 2(1+\tau) {\alpha}_{\varepsilon}^2 r^2 \Big( ({\varepsilon}^2 \tau + \frac{\tilde{\lambda}}{\tilde{G}} +2) {\alpha}_{\varepsilon} r + ({\varepsilon}^2 \tau + \frac{\tilde{\lambda}}{\tilde{G}} + 6-4\varepsilon)    \Big)    e^{- {\alpha}_{\varepsilon} (r-a)}            &&  \\
 & \qquad \qquad    -4 \frac{{\alpha}_{\varepsilon}^2}{{\beta}^2} {(\varepsilon -1)}^2 \tau (3+ 3\beta r + {\beta}^2 r^2) e^{-\beta (r-a)}     \Bigg] \cos\theta  ,   && \label{eq26appC}
\end{aligned}
\ee
\be 
\begin{aligned} 
& {\tilde{\sigma}}_{r\theta}^{\mathrm{n}} =   \tilde{G} \Big( \frac{1}{r} \frac{\partial  \tilde{v}_{n,r}}{\partial\theta} + \frac{\partial  \tilde{v}_{n, \theta}}{\partial r} - \frac{ \tilde{v}_{n, \theta}}{r}  \Big)  &&  \\
& \qquad =   \frac{3a \,  \tilde{G} }{2 {\beta}^2 r^4} \frac{ \tilde{U} (s)}{{\Delta}_{\varepsilon}}   \Bigg[ 12 (1+\tau) (1+ a {\alpha}_{\varepsilon}) -2 a^2 {\beta}^2 {\tau}_{\varepsilon} (\varepsilon -1) (1+ a {\alpha}_{\varepsilon})  - a^2 {\alpha}_{\varepsilon}^2 {\tau}_{\varepsilon} (\varepsilon -1) (1+ a\beta + a^2 {\beta}^2)      &&  \\
& \qquad \qquad -2 \frac{{\alpha}_{\varepsilon}^2}{{\beta}^2} \Big[ \tau \Big( 3(\varepsilon -1)(1+ a\beta) + (\varepsilon -3) a^2 {\beta}^2 \Big) -2 a^2 {\beta}^2     \Big]   &&  \\
& \qquad\qquad   +\tau \frac{{\alpha}_{\varepsilon}^2}{{\beta}^2} (\varepsilon -1) e^{-\beta (r-a)} \Big(6+ 6\beta r + 3 {\beta}^2 r^2 + {\beta}^3 r^3 \Big) - 4(1+ \tau) e^{- {\alpha}_{\varepsilon} (r-a)} \Big(  3 + 3 {\alpha}_{\varepsilon} r + {\alpha}_{\varepsilon}^2 r^2  \Big)     \Bigg] \sin\theta  .   && \label{eq27appC}
\end{aligned}
\ee

\section{Short- and long-time responses} \label{appC}
The velocity fields of fluid and network for a rigid sphere moving with velocity $ \tilde{U} (s)$ in a poroelastic medium are given in Eqs. \eqref{eq17ab}.
The limiting values of all these quantities at $t=0$ and $t\to \infty$ can be computed analytically using the following limits in S-space: $f(t \to 0) =\lim_{s\to \infty} s\tilde{f}(s)$  and $f(t\to \infty) =\lim_{s\to 0} s\tilde{f}(s)$.
For a sphere moving with constant force ${\mathrm{F}}_{\circ}$,  the time-dependent velocity of the sphere would be $U (s)= \displaystyle{{R}^{-1}(s) \mathrm{F}(s)  \equiv M (s) \frac{\mathrm{F_{\circ}}}{s} }$.
The sphere starts to move with velocity 
\be 
 U (t \to 0) = \frac{\mathrm{F_{\circ}}}{6\pi \eta a} \Big(  \frac{9 {\phi}^2 + 6 a {\beta}_{\circ} \phi + 3 a^2 {\beta}_{\circ}^2 }{ 8 {\phi}^2 + 8 a {\beta}_{\circ} \phi + 3 a^2 {\beta}_{\circ}^2 }\Big) , \label{eq24}
\ee
and at $t \to \infty$ stops with total displacemet of
$ 
\mathrm{x} (t \to \infty) = \frac{\mathrm{F_{\circ}}}{6\pi \eta a} \tau \Big( \frac{5-6\nu }{4(1-\nu) }  \Big) 
$.
At $t \to \infty$ the velocity fields of the fluid and network is zero, and the fluid and network displacements are:
 \be 
 \begin{aligned}
 u_{r,f}^{\infty} &=  \frac{{\mathrm{F}}_{\circ} \tau }{6\pi \eta a} \Big( \frac{a }{4 {\beta}_{\circ}^2 r^3 (\nu -1)(\phi -1) } \Big) \Big( 3(1-2\nu )  (1+ a {\beta}_{\circ}) + a^2 {\beta}_{\circ}^2 (\phi -1)+ 6 {\beta}_{\circ}^2 r^2 (1-\nu )(1-\phi)   \label{eq26}  \qquad\qquad\qquad\qquad \\
 & \qquad\qquad \qquad\qquad\qquad\qquad      +3(2\nu -1)  (1+ {\beta}_{\circ} r) e^{- {\beta}_{\circ} (r-a)}  \Big)  , 
 \end{aligned} 
 \ee
 \be
  \begin{aligned}
 u_{\theta ,f}^{\infty} &=  \frac{{\mathrm{F}}_{\circ} \tau}{6\pi \eta a} \Big( \frac{a }{8 {\beta}_{\circ}^2 r^3 (\nu -1) (\phi -1)} \Big)  \Big( 3(1-2\nu)  (1+ a {\beta}_{\circ}) + a^2 {\beta}_{\circ}^2 (\phi -1) - 6 {\beta}_{\circ}^2 r^2 (1-\nu )   +3  \phi {\beta}_{\circ}^2 r^2 (3-4\nu)   \qquad    \label{eq27}  \\
 & \qquad\qquad \qquad\qquad\qquad\qquad     +3(2\nu -1)   (1+ {\beta}_{\circ} r + {\beta}_{\circ}^2 r^2  ) e^{- {\beta}_{\circ} (r-a)}  \Big) ,
\end{aligned} 
\ee 
 \be 
u_{r,n}^{\infty} =  \frac{{\mathrm{F}}_{\circ} \tau }{6\pi \eta a} \Big( \frac{a^3 + 6 a r^2 (\nu -1)}{4 r^3 (\nu -1)}      \Big)  , \qquad\qquad\qquad\qquad  u_{\theta ,n}^{\infty} =  \frac{{\mathrm{F}}_{\circ} \tau }{6\pi \eta a} \Big( \frac{a^3 + 3 a r^2 (3-4\nu )}{8 r^3 (\nu -1)}      \Big) .   \label{eq28} 
 \ee

At $t=0$, the velocity of the fluid and network is Stokes velocity field, when the volume fraction of the network is zero, $\phi \to 0$.

\section{Relaxation dynamics of the fluid phase} \label{appD}
\begin{figure}
    \centering
            \subfigure[]{\includegraphics[width=0.485\textwidth , height=5.4cm]{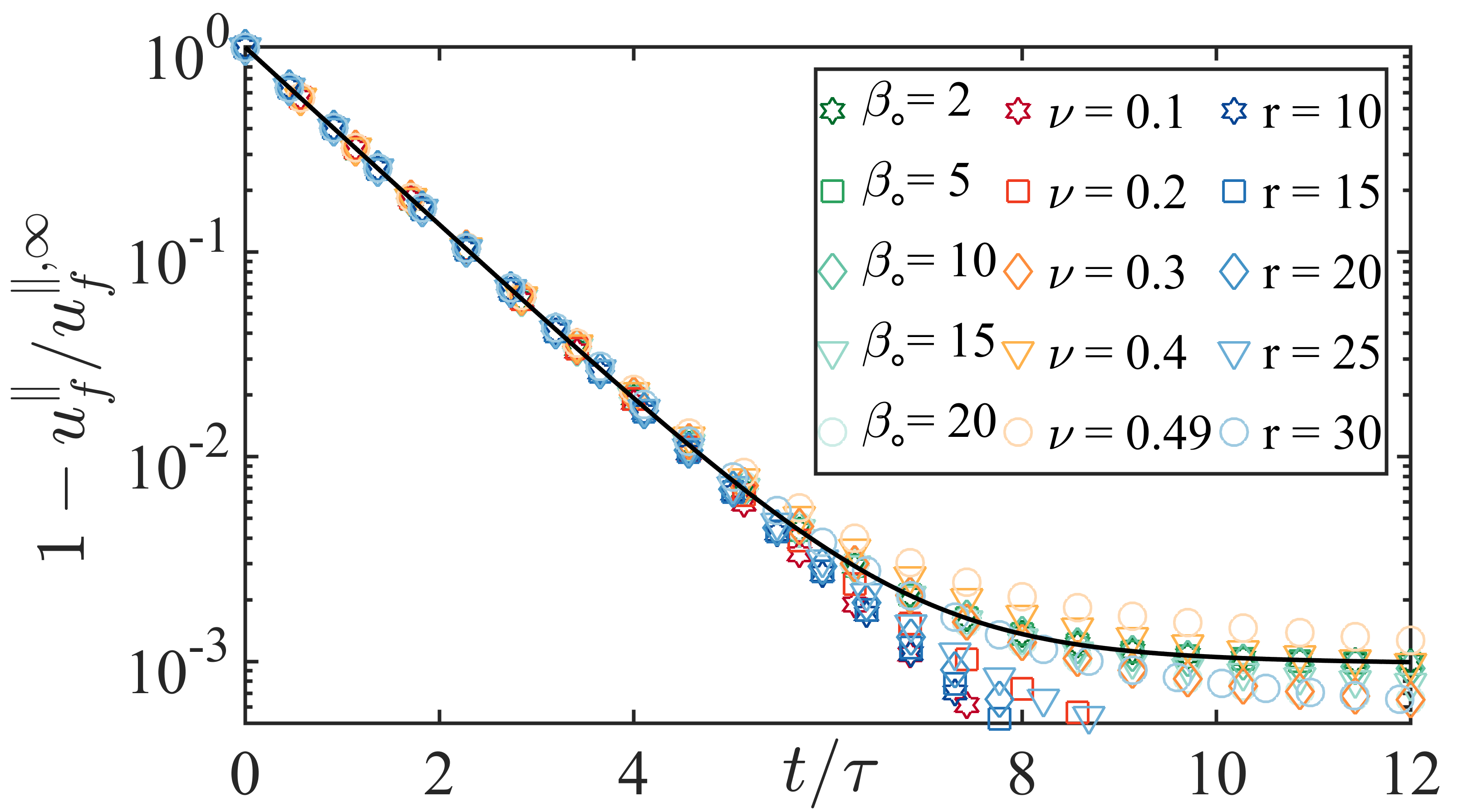}} 
        \subfigure[]{\includegraphics[width=0.485\textwidth , height=5.4cm]{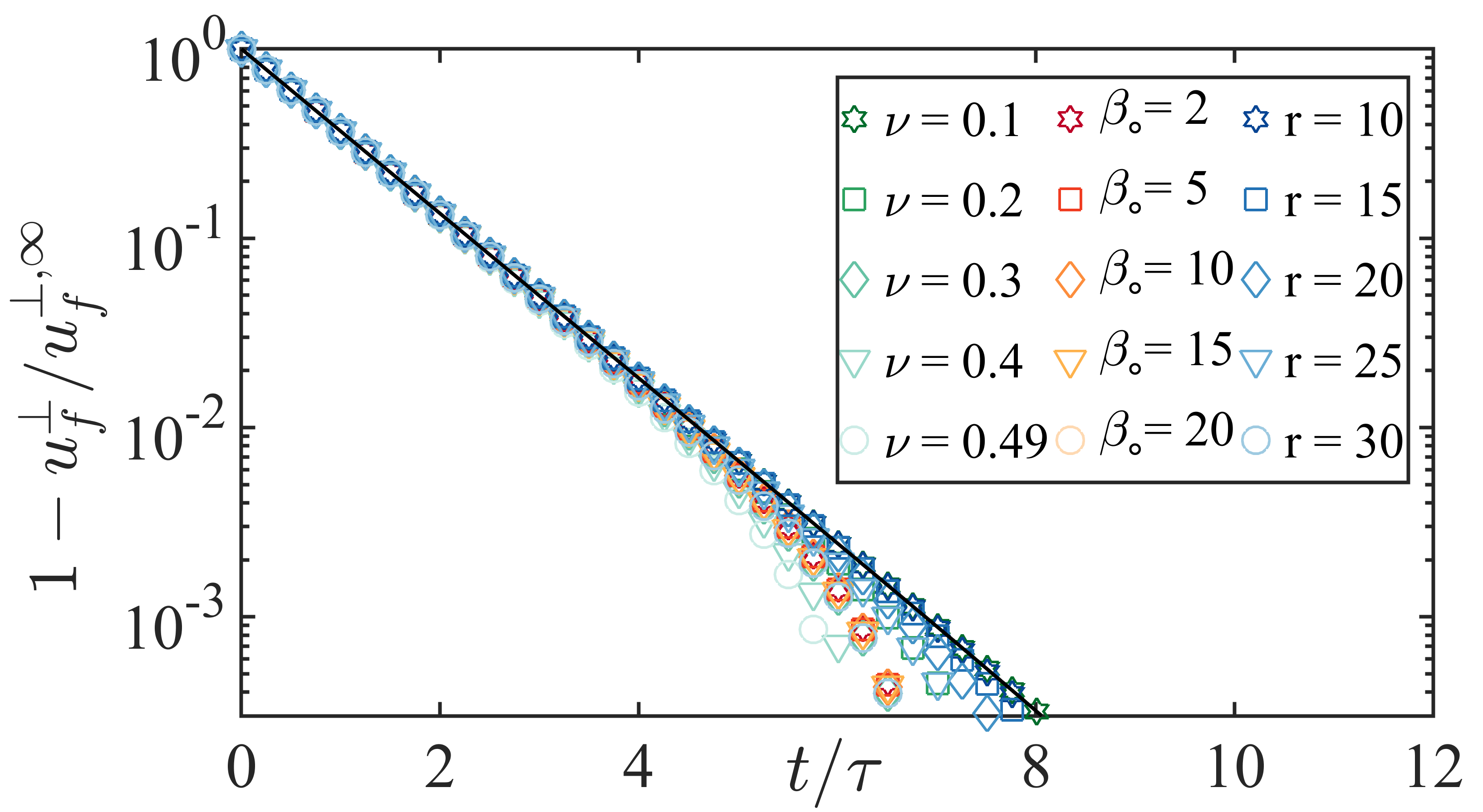}} 
    \caption{Relaxation of displacements in the direction parallel (a) and perpendicular direction (b) to the applied force for different values of permeability, Poisson ratio, $\nu$, and separation distances $r/a$. In all cases the relaxation is dominated (down to $1-u^{(\parallel,\perp)}_f/u_f^\infty \approx 0.001$) by the viscous relaxation timescale, $\tau$.  }
    \label{figS}
\end{figure}
Figure \ref{figS} shows the relaxation of the fluid displacement fields induced by the probe's motion in parallel (Fig.~\ref{figS}(a)) and perpendicular (Fig.~\ref{figS}(b)) directions for different 
values of permeability, $\beta_\circ$, separation distance, $r/a$, and Poisson ratios, $\nu$. 
As it can be seen, in all instances and in both parallel and perpendicular direction, the displacement relaxation is almost entirely (down to $1-u^{(\parallel,\perp)}_f/u_f^\infty \approx 0.001$) controlled by the relaxation timescale of viscous forces: $\tau$. 
\begin{figure}
    \centering
    \includegraphics[width=0.485\textwidth , height=5.4cm]{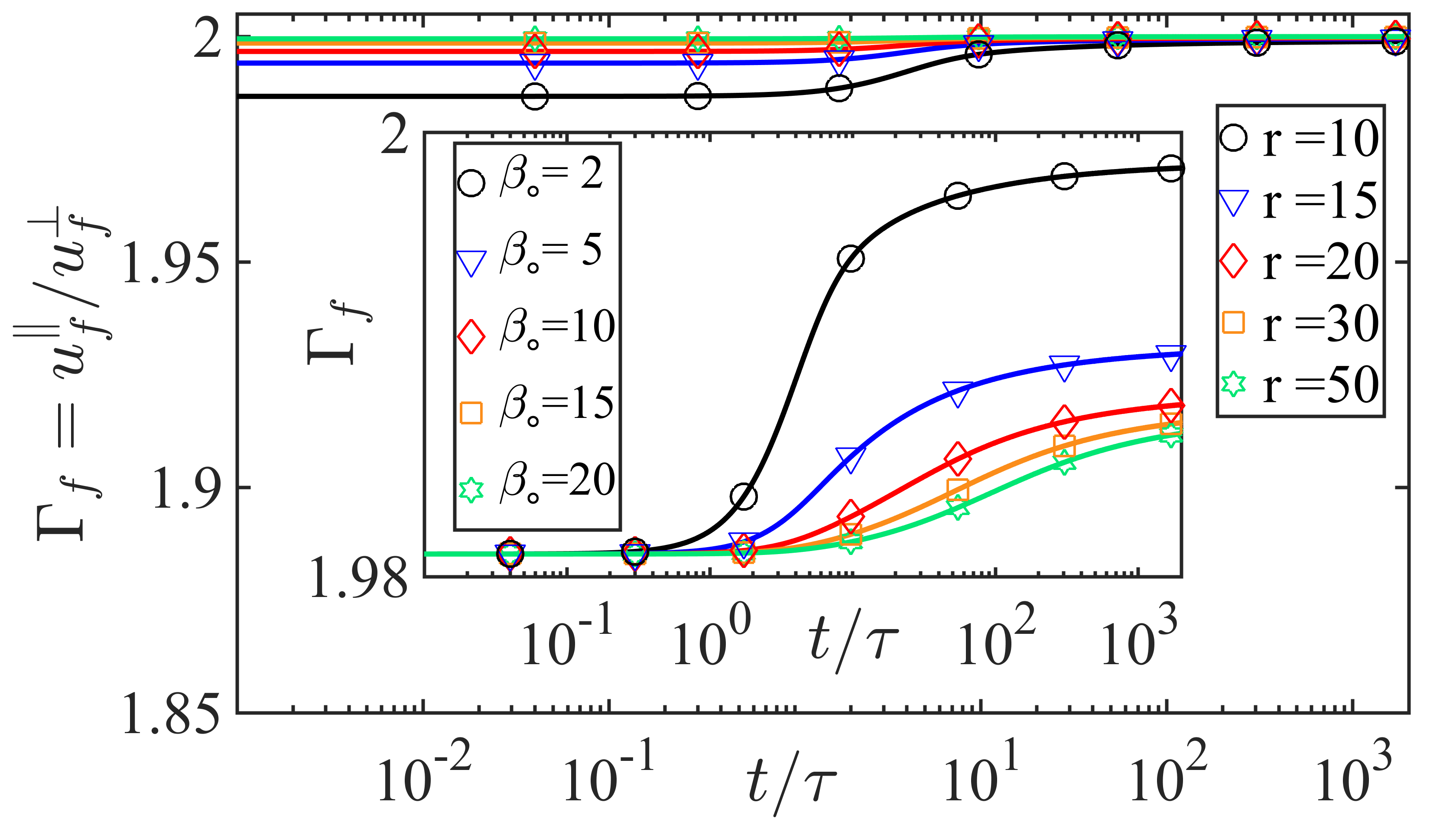}
    \caption{The ratio of parallel to perpendicular fluid displacements, $\Gamma_f=u^\parallel_f/u^\perp_f$, as a function of $t/\tau$. \textit{Main}: Different values of  $t/\tau$, taking $\beta_\circ=2$, and $\nu=0.3$. \textit{Inset}: Different permeabilities, for $r/a=10$, and $\nu=0.3$. }
    \label{fig:S2}
\end{figure}
Figure \ref{fig:S2} shows the ratio of parallel to perpendicular fluid displacements, $\Gamma_f=u^\parallel_f/u^\perp_f$, as a function of $t/\tau$. Main figure: for different values of  $t/\tau$, taking $\beta_\circ=2$, and $\nu=0.3$. Inset: Different permeabilities, and $r=10$, $\nu=0.3$. We see that this ratio remains nearly 2 (within 2\% error) , indicating a VE response.
\\
\\
\bibliographystyle{unsrt}
\bibliography{ENjfm}

\end{document}